\documentclass[aps, prd, 12pt, eqsecnum, tightenlines, notitlepage, nofootinbib, floatfix]{revtex4-1}

\PassOptionsToPackage{unicode}{hyperref}

\usepackage{hyperref, graphicx, slashed, xcolor, amsmath}

\allowdisplaybreaks

\definecolor{red}{rgb}{1,0,0}

\newcommand{\beq}{\begin{equation}}
\newcommand{\eeq}{\end{equation}}
\newcommand{\bea}{\begin{eqnarray}}
\newcommand{\eea}{\end{eqnarray}}
\newcommand{\nn}{\nonumber\\}

\begin{document}

\normalsize
\vspace{-0.5cm}
\begin{flushright}
CALT-TH-2017-033\\
\vspace{1cm}
\end{flushright}


\title{Quasi-Single Field Inflation in the non-perturbative regime}

\author{Haipeng An, Michael McAneny, Alexander K. Ridgway and Mark B. Wise}
\affiliation{Walter Burke Institute for Theoretical Physics,
California Institute of Technology, Pasadena, CA 91125}

\begin{abstract}

In quasi-single field inflation there are  massive fields that interact with the inflaton field. If these other fields are not much heavier than the Hubble constant during inflation ($H$) these interactions  can lead to important consequences for the cosmological energy density perturbations. The simplest model of this type has a real scalar inflaton field  that interacts with another real scalar $S$ (with mass $m$). In this model there is a mixing term of the form $\mu  {\dot \pi}  S$, where $\pi$ is the Goldstone fluctuation that is associated with the breaking of time translation invariance by the time evolution of the inflaton field during the inflationary era. In this paper we study this model in the  region $(\mu/H )^2 +(m/H)^2 >9/4$ and $m/H \sim {\cal O}(1)$ or less. For a large part of the parameter space in this region standard perturbative methods are not applicable. Using numerical and analytic methods we study how large $\mu/H$ has to be for the large $\mu/H$ effective field theory approach to be applicable.
\end{abstract}

\maketitle

\section{Introduction}
There is very strong evidence that the universe was once in a radiation dominated era followed by a matter dominated  era. Today the universe is dominated by vacuum energy density and we are entering an inflationary era where the scale factor $a(t) \propto e^{H_0t}$, with $H_0$ near the Hubble constant today. It is widely believed that at very early times there was another inflationary era where the energy density was dominated by false vacuum energy giving rise to a Robertson Walker scale factor with time dependence $a(t) \propto e^{Ht}$, where $H$ is the Hubble constant during that inflationary era~\cite{SKS, GuthLindeAlbrechtSteinhardt}. After more than about 60 e-folds, this inflationary era ends and the universe reheats to a radiation dominated (Robertson Walker) Universe. If this is the case then the horizon and flatness problems~\cite{GuthLindeAlbrechtSteinhardt} can be solved and in addition there is an attractive mechanism based on quantum fluctuations for generating  density perturbations with wavelengths that were once outside the horizon~\cite{mshsgpbst} (see Ref.~\cite{Baumann:2009ds} for a review of inflation).  It has been argued that it requires tuning to enter the inflationary era~\cite{PGT} (see however \cite{Bastero-Gil:2016mrl}) and furthermore that there are issues with its predictability~\cite{SVG} (see also~\cite{Ijjas:2013vea} for a recent discussion of these issues). Nevertheless, because of the simplicity of the  dynamics of the inflationary universe paradigm and the ability within it to do explicit calculations  of the properties of the cosmological energy density perturbations~\cite{mshsgpbst} and primordial gravitational waves~\cite{GSRAW}, it seems worth studying particular inflationary models in some detail.

The simplest inflationary model is standard slow roll inflation with only a single real scalar field, the inflaton  $\phi(x)$. It is conventional to work in a gauge where fluctuations in the inflaton field about the classical slow roll solution $\phi_0(t)$ vanish. Then using the St$\ddot{\rm u}$ckelberg trick the curvature fluctuations that are constant outside the horizon and become the density perturbations when they reenter the horizon (in the radiation and matter dominated eras) arise from quantum correlations in the Goldstone mode $\pi(x)$ calculated during the de-Sitter inflationary era\footnote{The effective field theory formulation for inflation~\cite{Cheung:2007st} provides an elegant method to compute correlations of $\pi$ in a model independent fashion.}. In this model non gaussianities in cosmological density correlations arise because of connected higher point correlations of $\pi$, but they are very small~\cite{Maldacena:2002vr}. 

Larger non-gaussianities can be achieved if there are other fields with masses around or less than the inflationary Hubble constant\footnote{There are other ways to have large non-gaussianities.  For example, DBI inflation~\cite{Alishahiha:2004eh}. For an early example of another type, see~\cite{AGW}.}, that couple to $\pi$ (see Ref.~\cite{Chen:2010xka} for a review). In quasi-single field inflation these extra fields do not directly influence the classical evolution of the inflaton field but impact the cosmological density perturbations since they couple to the inflaton as ``virtual particles'' and hence affect the the correlations of  $\pi$~\cite{Chen:2009zp}.  To simplify matters we will assume an approximate shift symmetry on the inflaton field, $\phi(x) \rightarrow \phi(x) +c$ (where $c$ is a constant) that is only broken by the potential, $V_{\phi}$, for $\phi$. Furthermore, we assume an unbroken discrete symmetry, $\phi(x) \rightarrow -\phi(x)$.  The simplest quasi-single field model introduced by Chen and Wang~\cite{Chen:2010xka} has  a single additional (beyond the inflaton) real scalar field $S$.  The  Lagrange density in this model contains an unusual kinetic mixing of the form $\mu  {\dot \pi}  S$ . 

This model has been extensively studied in the perturbative region\footnote{By perturbation theory we mean a series expansion in $\mu/H$.} where $\mu /H  \ll 1$~\cite{Chen:2009zp,Baumann:2011nk,Assassi:2012zq,Noumi:2012vr,Arkani-Hamed:2015bza,Tolley:2009fg,Baumann:2011su,Achucarro:2012sm,Achucarro:2012yr,Jiang:2017nou}.  In the non-perturbative region where $\mu/H \gg1$,  an elegant effective field theory formulation has been derived by Baumann and Green \cite{Baumann:2011su}, and by Gwyn, Palma, Sakellariadou, and Sypsas \cite{Gwyn:2012mw}.  The curvature perturbation power spectrum and a contribution to its bispectrum have been calculated using this formulation.  It has been studied numericaly in \cite{Assassi:2013gxa} for other regions of the parameter space.  

Throughout this paper we treat $\mu$ as a constant independent of time. There has been a study of the case where $\mu$ changes suddenly with time, becoming large momentarily~\cite{Shiu:2011qw}.

 In this paper we focus on the region of parameter space where $(\mu/H)^2 +(m/H)^2  >9/4$ and $m/H \sim {\cal O}(1)$ or less (recall $m$ is the mass term for $S$).  In this region, non-gaussianities have an interesting oscillatory behavior~\cite{Arkani-Hamed:2015bza}.  We use numerical non-perturbative  methods similar to those developed in \cite{Assassi:2013gxa} and the effective field theory for large $\mu/H$ to study the model in this region of parameter space.  We  study how large $\mu/H$ must be for the effective field theory method to be quantitatively correct. In addition we derive the $n_S$, $r$ plot for the model with inflaton potential $V_{\phi}= m_{\phi}^2 \phi^2/2$ and derive the limit on $\mu/H$ and the $S$ potential parameter $V_S'''$ from Planck limits on non-gaussianity.  
  
 In section~\ref{The Model} we discuss the Lagrange density of the model we use in detail. Section \ref{Free Field Theory in Flat Space-time}  reviews  quantization of the free part of the Lagrange density in flat space-time. Even this theory is non-trivial because of the unusual Lorentz non-invariant kinetic mixing between the Goldstone field $\pi$ and the excitations of the massive scalar $S$. The massless mode has an unusual energy momentum relation that, for momentum in the range $m \ll q \ll \mu$, has a non-relativistic flavor, $E_q=q^2/ \mu$~\cite{Baumann:2011su}.  The other mode is heavy with a mass $\sqrt{\mu^2 +m^2}$. The fact that this mode's mass does not go to zero as $m \rightarrow 0$ is what regularizes the divergences that occur at $m=0$ when one treats $\mu$ perturbatively.  

Quantization of the free field theory in de-Sitter space-time is discussed in section \ref{Free Field Theory in de-Sitter Space time}.  In de-Sitter space-time  a mode's physical momentum $q$ evolves with time. At early times modes have wavelengths much less than the horizon $1/H$ but at later times the wavelengths get red-shifted outside the horizon.  The mode functions are calculated non-perturbatively by numerically solving the differential equations they satisfy in the region of parameter space, $(\mu/H)^2 +(m/H)^2  >9/4$ and $m/H \sim {\cal O}(1)$ or less. Quantum fluctuations in the  field $S$ fall off rapidly for wavelengths outside the horizon and it is the quantum fluctuations in the field $\pi$ that determine the curvature and density fluctuations just as in standard slow roll single field inflation. Nevertheless, these quantum fluctuations are influenced by $\pi$'s couplings to $S$.  

In section~\ref{sec:freetheory} we analyze (in the non-perturbative regime) the curvature perturbation power spectrum in this model focusing on the transition between the perturbative regime and the regime where the effective theory applies.  Section~\ref{sec:observable} derives the $n_S, r$ plot in this theory for the simple inflaton potential $V_{\phi}=m_{\phi}^2\phi^2/2$.

Non-gaussianities are discussed in Sec~\ref{sec:nonGaussian}. We calculate the bispectrum in the the equilateral and squeezed configurations in the non-perturbative region numerically. In the large $\mu$ region we show that the numerical results agree with the results from the effective theory. We derive the constraints on $\mu/H$ and the $S$ potential parameter $V_S'''$ from Planck limits on non-gaussianity.

In section~\ref{sec:effetivetheory} we review the derivation of the effective field theory for large $\mu/H$ and the derivation of the power spectrum using it. We then compute the bispectrum in this effective field theory including a contribution from the potential for $S$ that was not previously presented in the literature.

Our conclusions are summarized in Sec.~\ref{sec:conclusion}.

\section{The Model} 
\label{The Model}

The simplest quasi-single field inflation model  has a real scalar inflaton field $\phi$ that interacts with another real scalar field $S$. We impose a $\phi \rightarrow -\phi$ symmetry and an approximate shift symmetry $\phi \rightarrow \phi+c$, where $c$ is a constant. The shift symmetry is only broken by the inflaton potential $V_{\phi}(\phi)$. The Lagrangian we use has the form

\begin{equation} 
{\cal L}={\cal L}_{\phi}+{\cal L}_{S}+{\cal L}_{\rm int}
\end{equation}
where
\begin{equation}\label{eq:Lorgininal}
{\cal L}_{\phi}= {1 \over 2}g^{\mu \nu} \partial_{\mu}\phi \partial_{\nu} \phi -V_{\phi}(\phi),~~~{\cal L}_{\rm S}= {1 \over 2}g^{\mu \nu} \partial_{\mu}S \partial_{\nu} S-V_{S}(S)
\end{equation}
Interactions between the inflaton $\phi$ and the massive field $S$ first occur at dimension 5 and  if we neglect operators with  dimension higher than this the interaction Lagrangian is
\begin{equation}\label{eq:int}
{\cal L}_{\rm int}={1 \over \Lambda} g^{\mu \nu} \partial_{\mu}\phi \partial_{\nu} \phi S.
\end{equation}
One natural choice for the mass scale $\Lambda$ is the Planck mass. This higher dimensional operator would then arise from the transition from the theory of quantum gravity to a quantum field theory. In this case the non-gaussianities are very small. However, another possibility is that there is physics at a scale $\Lambda$ that is large compared to the Hubble constant during inflation but well below the Planck scale. Integrating out this physics can give rise to such an operator. 

We  work in a gauge where the inflaton field is only a function of time, $\phi(x)=\phi_0(t)$ and take the background metric to have the form,  $ds^2=dt^2-a(t)^2d{\bf x}^2$, with the scale factor $a(t)=e^{Ht}$. The Goldstone boson associated with the time translation invariance breaking by the classical evolution  $\phi_0(t) $ is denoted by $\pi(x)$. The curvature perturbation is proportional to this field, $\zeta=-H \pi$.  We expand $S$ about a background classical value $S(x)=S_0 +s(x)$ and assume that the background solution $S_0$ is independent of time. This assumption is consistent with the dynamical equations of evolution for the fields provided we neglect second time derivatives of $\phi_0(t)$.    With those assumptions $\phi_0(t) $ and $S_0$ satisfy,
\begin{equation}
\left(1+ {2S_0 \over \Lambda} \right)3 H {\dot \phi}_0+{d V_{\phi}(\phi_0) \over d\phi_0} =0,
\end{equation}
and
\begin{equation}
{{\dot \phi}_0^2 \over \Lambda}-{d V_{S}(S_0) \over dS_0} =0.
\end{equation}
The dynamics for the fluctuations $\pi(x)$ and $s(x)$ are controlled by the Lagrange density,
\begin{equation}
{\cal L}= {\cal L}_0+{\cal L}_{\rm int}
\end{equation}
where the free part of the Lagrange density for the fields $\pi$ and $s$ is,
\begin{equation}
{\cal L}_0={1 \over 2}\dot\phi_0^2\left(1+{ 2S_0 \over \Lambda} \right) \left({\dot \pi}^2- {1 \over a^2} {\nabla \pi} \cdot  {\nabla \pi} \right)+{1 \over 2}\left( {\dot s}^2 - {1 \over a^2} \nabla s \cdot \nabla s  -m^2  s^2 \right) +{2 \over \Lambda} {\dot \phi}_0^2{\dot \pi} s
\end{equation}
where $m^2=V''(S_0)$. Throughout this paper we assume that the mass parameter $m$ for the additional scalar  $s$ is of order the Hubble constant during inflation or smaller.

The interaction part of the Lagrange density is
\begin{equation}
\label{interaction Lagrangian}
{\cal L}_{\rm int}= {{\dot \phi_0}^2 \over \Lambda} \left({\dot \pi}^2- {1 \over a^2} {\nabla \pi} \cdot  {\nabla \pi} \right)s+\left(  {\dot \pi} +{ {\dot \pi}^2 \over 2}\right) {\dot s}^2- { 1 \over 3!} V_S'''(S_0)s^3-{ 1\over 4!} V_S''''(S_0)s^4+\ldots 
\end{equation}
It is convenient to introduce a rescaled $\pi$ that has a properly normalized kinetic term,
\begin{equation}
{\tilde \pi}=\sqrt{{\dot \phi_0}^2(1+2S_0/\Lambda)} \pi =|{\dot {\tilde \phi}_0}|\pi
\end{equation}
where,
\begin{equation}
{\tilde \phi}_0=\sqrt{(1+2S_0/\Lambda)} \phi_0. 
\end{equation}
In terms of these rescaled fields the gravitational curvature perturbation becomes, 
\begin{equation}
\zeta=-(H / |{\dot{\tilde \phi}}_0|)\tilde \pi. 
\end{equation}
The free and interacting Lagrange densities, after introducing a redefined scale  ${\tilde \Lambda}=(1+2S_0/\Lambda)\Lambda$, are
\begin{equation}
\label{fft}
{\cal L}_0={1 \over 2} \left({\dot{\tilde  \pi}}^2- {1 \over a^2} {\nabla {\tilde \pi}} \cdot  {\nabla {\tilde \pi}} \right)+{1 \over 2}\left( {\dot s}^2 - {1 \over a^2} \nabla s \cdot \nabla s  -m^2  s^2 \right) +\mu {\dot {\tilde \pi} }s
\end{equation}
and
\begin{equation}
\label{intlagrange}
{\cal L}_{\rm int}  =  {1 \over {\tilde \Lambda}} \left({\dot {\tilde \pi}}^2- {1 \over a^2} {\nabla {\tilde \pi}} \cdot  {\nabla \tilde \pi} \right)s - { 1 \over 3!} V_S'''(S_0)s^3 + \ldots .
\end{equation}
In eq.~(\ref{fft}) we have introduced
\begin{equation}\label{eq:mu}
 \mu=2{\dot {\tilde \phi}_0}/{{\tilde \Lambda}}. 
 \end{equation}
and in eq.~(\ref{intlagrange}) only explicitly kept those terms that play a role in the calculations performed in this paper. In the following sections  we will drop the tilde on the Goldstone field ${\tilde \pi}$ to simplify the notation.  Moreover, we adopt sign conventions for  $\phi$ and $S$ so that $\dot\phi_0$ and $\mu$ are positive.

As mentioned in the introduction the purpose of this paper is to study this model in the region of parameter space  where $(\mu^2+ m^2)^{1/2}/H > 3/2$ and $m \sim {\cal O}(H)$ or smaller. Some of this region, {\it i.e.} where $\mu/H$ is small or very large have been previously studied. We will compare with those results to find out how small  and how large $\mu/H$ has to be for the approximate methods used in those regions to be accurate. 

First let's imagine that $S_0$=0. This can always be arranged by tuning the linear term in the potential $V_S(S)$ to cancel the linear term in $S$ from the $1/\Lambda$  interaction term. 
Then $\mu/H= (2 {\dot \phi_0}/H^2) (H/\Lambda)$. The measured power spectrum for the curvature perturbations implies that  ${\dot \phi_0}/H^2$ is very large so even for small $H/\Lambda$ one can achieve large values for $\mu/H$. 

Next we allow a non zero $S_0$ but simplify the potential so it contains no terms with more than two powers of $S$, explicitly  $V_{S}= V_S'S+ m^2 S^2/2$.  In this case  $\mu/H$ can be written as,
\beq
\frac{\mu}{H} = \frac{2 \dot \phi_0 / \Lambda H}{\left[1 + 2\frac{(\dot\phi_0/\Lambda H)^2 - V'_S/H^2\Lambda}{m^2/H^2}\right]^{1/2}} \ .
\eeq
Therefore, without tuning the tadpole in $V_S$ to cancel  $\dot \phi_0^2 S/\Lambda$,  it is not possible to have  the mass parameter $m$ of order the Hubble constant (or smaller) and $\mu/H$ large. Nonetheless it seems worth studying this region of parameter space since there are some novel features that arise there.

Naive dimensional analysis suggests that higher dimension operators that couple derivatives of $\phi$ to a single $S$ are smaller than the dimension 5 operator we kept provided ${\dot \phi_0}/\Lambda^2 =(\mu/H)^2 (H^2 /{\dot\phi_0}) <1$. The higher powers of $S$ will be small if in addition $S_0/\Lambda <1$. Since the measured amplitude of the density perturbations implies that $H^2 /{\dot\phi_0} $  is quite small the ratio $\mu/H$ can be large in the region of parameter space where the operator expansion in powers of ${1/\Lambda}$ is justified. Indeed, comparing the calculated power spectrum at  large $\mu/H$  given in ~(\ref{eq:power}) with it's measured value, the upper limit for $\mu/H$ for  power counting in the $1/\Lambda$ expansion to be valid is $\mu / H \lesssim 300$. Of course, this is just a naturalness constraint and can be violated without the model being inconsistent.

\section{Free Field Theory in Flat Space-time} 
\label{Free Field Theory in Flat Space-time} 

In this section we review, for pedagogical reasons, quantization in flat space-time of the free field theory with Lagrange density in eq.~(\ref{fft}). The results presented here have,  by in large, been noted previously in~\cite{Baumann:2011su,Gwyn:2012mw}.     

Dropping the tildes and setting $a(t)=1$ the Lagrange density in eq.~(\ref{fft})  becomes,
\begin{equation}\label{eq:Lflat}
{\cal L}_0={1 \over 2} \left({\dot  \pi}^2- {\nabla  \pi} \cdot  {\nabla  \pi} \right)+{1 \over 2}\left( {\dot s}^2 - \nabla s \cdot \nabla s  -m^2  s^2 \right) +{  \mu}{\dot  \pi }s.
\end{equation}
This corresponds to normal kinetic terms for two real scalar fields but with an unusual Lorentz non-invariant kinetic mixing. The Lagrange density has the shift symmetry ${\pi} \rightarrow { \pi} +c$ for the Goldstone field $\pi$. 

The classical equations of motion for the fields $\pi$ and  $s$ are,
\begin{equation}
{\ddot \pi} -{\nabla^2  \pi} +\mu{\dot {s}}=0
\end{equation}
and
\begin{equation}\label{sflat}
{\ddot s}-{\nabla}^2 s +{m^2}s-\mu {\dot \pi}=0
\end{equation}
Quantization proceeds by expanding the fields in modes,
\begin{equation}
\label{piflat}
 \pi({\bf x},t)=\int {d^3 q \over (2 \pi)^3} \left( a^{(1)}({\bf q})  \pi_{q}^{(1)}(t) e^{i{\bf q}\cdot {\bf x}}+a^{(2)}({\bf q}) \pi_{q}^{(2)}(t)e^{i{\bf q} \cdot {\bf x}}  +{\rm h.c.}  \right)
 \end{equation}
 and
\begin{equation}
 s({\bf x},t)=\int {d^3 q \over (2 \pi)^3} \left( a^{(1)}({\bf q})  s_{q}^{(1)}(t) e^{i{\bf q}\cdot {\bf x}}+a^{(2)}({\bf q}) s_{q}^{(2)}(t)e^{i{\bf q} \cdot {\bf x}}  +{\rm h.c.}  \right)
\end{equation}
 The annihilation operators $a^{(1,2)}({\bf q})$ and creation operators $a^{(1,2)}({\bf q})^{\dagger}$ satisfy the usual commutation relations\footnote{More explicitly the non-zero commutators are: $[a^{(1)}({\bf q}),a^{(1)}({\bf q}')^{\dagger}]=(2 \pi)^3\delta^3({\bf q}-{\bf q'})$ and $[a^{(2)}({\bf q}),a^{(2)}({\bf q}')^{\dagger}]=(2 \pi)^3\delta^3({\bf q}-{\bf q'})$}. The time dependence of the mode functions $\pi_{q}^{(1,2)}(t)$ and  $s_{q}^{(1,2)}(t)$  are determined by solving the classical equations of motion and their normalization is fixed by the canonical commutation relations of the fields with  their canonical momenta. A  difference from the usual case where there is no Lorentz non-invariant mixing is that the canonical momentum for the field $\pi$ is not ${\dot \pi}$ but rather ${\dot \pi}+\mu s$. So ${\dot \pi}$ and ${\dot s}$ don't commute at equal time but rather satisfy $[{\dot \pi}({\bf x},t),{\dot s}({\bf x}',t)]=- i\mu\delta^3({\bf x}-{\bf x}')$. 
 
 The time dependence of the modes has the usual exponential form $\pi_q^{(1,2)}(t) \propto \exp(-iE_q^{(1,2)}t)$, $s_q^{(1,2)}(t) \propto \exp(-iE_q^{(1,2)}t)$ . The dispersion relations for the energies is determined by solving the classical equations of motion. This yields,
 \begin{equation}
 E_q^{(1,2)}=\left[q^2 +{m^2 +\mu^2 \over 2} \pm {1 \over 2}\left((m^2 +\mu^2)^2+4q^2\mu^2 \right)^{1/2}\right] ^{1/2}\ ,
 \end{equation}
which is a massless mode  that we label by (1) corresponding to the minus sign and a massive mode that we label by (2) corresponding to the plus sign. The mass of mode (2) is $\sqrt{m^2+\mu^2}$. Because this mode remains massive even when $m=0$ there will be no divergences in our calculations in de-Sitter space.
 
 
We now focus on the large mixing region of parameter space, $\mu \gg  q, m$. As discussed in the literature~\cite{Baumann:2011su} the dispersion relations of the two modes can be written as
\begin{equation}
\label{eq:flatdispersion}
E_q^{(1)} \simeq{q\sqrt{q^2 +m^2} \over  \mu},~~~E_q^{(2)} \simeq \mu.
\end{equation}
The $(1)$ mode is massless but for $q$ much larger than $m$ the energy grows not linearly with $q$ but rather quadratically (like a non relativistic particle). The other mode is massive with mass $ \mu$.  For very small momentum, $q  \ll m$, the massive scalar $s$ only contains the massive $(2)$ mode, {\it i.e.}, $|s^{(1)}_q(t)/ s^{(2)}_q(t)| \rightarrow 0$ as $ q \rightarrow 0$.  On the other hand the Goldstone field $\pi$ contains equal amounts of the $(1)$ and $(2)$ modes. 

The infrared, $q \rightarrow 0$,  behavior of  the mode function $s^{(1,2)}_q$  changes in the special case that $m=0$. Then integrating-by-parts, the kinetic mixing term in Eq.~(\ref{eq:Lflat}) can be recast as $-\mu \pi\dot s$, and so it is clear that there is also a  shift symmetry in $s$. For $m=0$ the scalar field $s$ also contains equal amounts of the two modes.

Since for large $\mu$  the second mode is heavy  it is appropriate for the physics at low momentum $q\ll\mu$ to integrate it out from the theory and write an effective Lagrange density in terms of a single massless field. For the light mode a time derivative gives factors of $1/\mu$ and  (for $m \ne 0$) at very large $\mu$ the $s$ field contains only a small amount of that massless mode. Hence  eq.(\ref{sflat}) imples that,
\begin{equation}
s \simeq \left( {\mu \over m^2-\nabla^2} \right) {\dot \pi}
\end{equation}
Putting this into the Lagrange density in eq. (\ref{eq:Lflat}) and dropping terms suppressed by powers of $1/\mu$ (recall a time derivative on $\pi$ is suppressed by $1/\mu$) yields the effective Lagrange density for the massless mode,
\begin{equation}
{\cal L}_{\rm eff}= {1 \over 2} \left( {\mu^2 \over  m^2-\nabla^2} \right) \dot \pi^2 -{1 \over 2} {\nabla \pi} \cdot {\nabla \pi}
\end{equation}
which yields the dispersion relation for the massless mode given in eq.~(\ref{eq:flatdispersion}). 

In the next section we perform the quantization in curved de-Sitter space-time (with Hubble constant $H$). Then the physics of the massless $(1)$ mode should be similar to that in flat space-time when the momentum and energy for that mode are large compared to $H$  {\it i.e.},  $q>  H$ and $E_q^{(1)}  >  H$. In the flat space-time large $\mu$ discussion we assumed $q< \mu$. The energy condition $E_q^{(1)} >  H$ implies $q$ must also satisfy $q  > \sqrt{\mu H}$ in order for our de-Sitter space-time computations to resemble the flat space-time  large $\mu$ case discussed in this subsection.  

\section{Free Field Theory in de-Sitter Space time}\label{Free Field Theory in de-Sitter Space time}
\label{sec:freetheory}

Introducing conformal time, $\tau= -e^{-Ht}/H$, and including the measure factor $\sqrt{-g}$ in the Lagrange density so that the action is equal to $\int d^3x  d\tau  {\cal L}$ we have
\begin{equation}\label{eq:Lagrange}
{\cal L}_0=\frac{1}{2H^2\tau^2} \left( \left(\partial_{\tau}\pi\right)^2 - \nabla\pi\cdot\nabla\pi +  \left(\partial_{\tau}s\right)^2 - \frac{m^2}{H^2 \tau^2} {s^2} - \nabla s\cdot \nabla s  - \frac{2\mu }{H \tau}  s{ \partial_{\tau} \pi } \right) \ .
\end{equation}
As in flat space we expand the quantum fields in terms of creation and annihilation operators. Introducing $\eta=k \tau$ we write,
\begin{equation}\label{eq:mode}
 \pi({\bf x},\tau)=\int {d^3 k \over (2 \pi)^3} \left( a^{(1)}({\bf k})  \pi_{k}^{(1)}(\eta) e^{i{\bf k}\cdot {\bf x}}+a^{(2)}({\bf k}) \pi_{k}^{(2)}(\eta)e^{i{\bf k} \cdot {\bf x}}  +{\rm h.c.}  \right)
\end{equation}
and
\begin{equation}
 {s}({\bf x},\tau)=\int {d^3 k \over (2 \pi)^3} \left( a^{(1)}({\bf k})  {s}_{k}^{(1)}(\eta) e^{i{\bf k}\cdot {\bf x}}+a^{(2)}({\bf k}) {s}_{k}^{(2)}(\eta)e^{i{\bf k} \cdot {\bf x}}  +{\rm h.c.}  \right)
 \end{equation}
 The mode functions obey the classical equations of motion,
\begin{equation}\label{eq:diff1}
\pi''_k - \frac{2\pi'_k}{\eta} + \pi_k - \frac{\mu}{H} \left(  \frac{s'_k}{\eta} - \frac{3s_k}{\eta^2}\right) = 0 \ 
 \end{equation}
and
\begin{equation}\label{eq:diff2}
s''_k-\frac{2s'_k}{\eta} + \left(1+\frac{m^2}{H^2\eta^2}\right)s_k + \frac{\mu}{H} \frac{\pi'_k}{\eta} = 0 \ ,
\end{equation}
 where  a `` $'$ '' represents an $\eta$ derivative. 
 
 \subsection{Numerical results}
 
In the mode expansion for for the fields ${s}$ and $\pi$, $k$ is the magnitude of the comoving wavevector. The physical wavevector has magnitude $q=k/a= -H  \eta$. Hence the condition that a mode have wavelength  well within  the de-Sitter horizon $1/H$ is $q/H \gg 1$ which is equivalent to $-\eta  \ll 1$. At fixed $k$ as time evolves a mode goes from physical wavelength well within the horizon to outside the horizon.

In the region well within the horizon, $-\eta\gg {\mu/H}$ and $- \eta \gg 1$, the differential equations (\ref{eq:diff1}) and (\ref{eq:diff2})  simplify to 
\bea
 \pi''_k + \pi_k &=& 0 \ ,\nn 
 {s}''_k + {s}_k &=& 0 \ .\nn
\eea
Here we suppressed the superscripts $(1,2)$ that label mode type. 
The leading behavior of the mode functions is
\bea
\pi_k \sim {s}_k \sim e^{-i\eta} 
\eea
and so it is convenient to represent the general solution in the region deeply inside the horizon as
\bea
\label{large tau limit}
\pi_k = A_k e^{- i\eta} \ , \;\; {s}_k = B_k e^{-i\eta} \ .
\eea
$A$ and $B$ are functions of $\eta$ with $ |A'/A|,  |B'/B| \ll 1$.  Substituting $\pi_k$ and ${s}_k$ back into (\ref{eq:diff1}) and (\ref{eq:diff2}) and keeping only the leading order terms in $\eta^{-1}$ we find
\bea
2 A' _k - \frac{2A}{\eta} - \frac{\mu}{H\eta} B_k &=&  0 \nn
2  B'_k - \frac{2B}{\eta} + \frac{\mu}{H\eta} A_k &=& 0 
\eea
which gives 
\beq
A _k\propto (-\eta)^{1\pm \frac{i\mu}{2H}} \ , \;\; B _k= \pm i A _k\ .
\eeq
Therefore, in this region the canonically normalized form of $\pi_k$ and $s_k$ can be written as
\beq\label{eq:UVboundary}
\pi_k^{(1,2)} = \frac{H}{\sqrt{4k^3}} e^{-i\eta } (-\eta)^{1\pm\frac{i\mu}{2H}} \ , \;\; {s}_k^{(1,2)} = \pm i \pi_k^{(1,2)} \ ,
\eeq
where the factor ${H}/\sqrt{4k^3}$ is determined by the canonical commutation relations. 

\begin{figure*}[tp]
\centering
\includegraphics[width=3.5in]{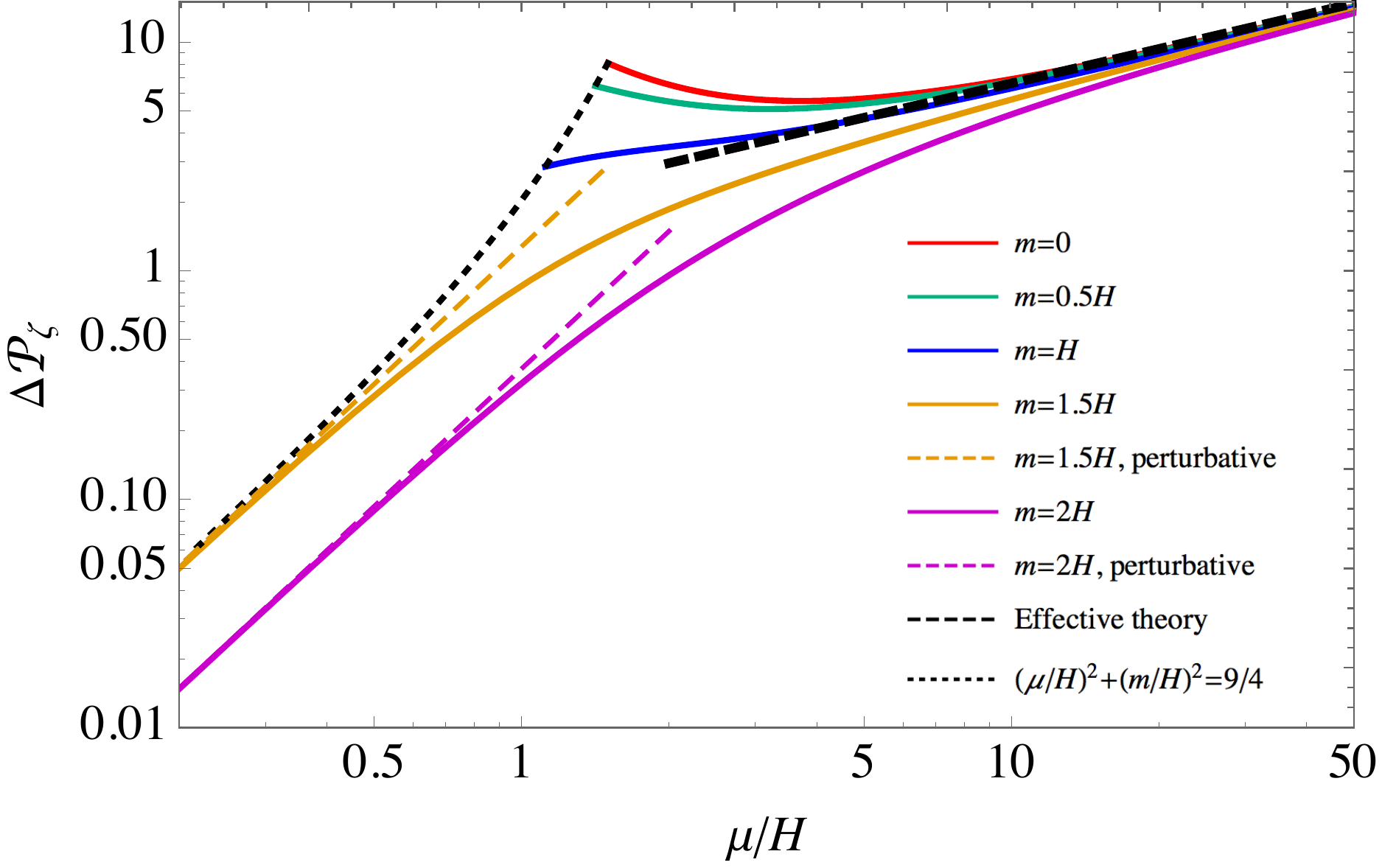}
\caption{The correction of the power spectrum of curvature perturbation ${\Delta\cal P}_\zeta$ in  units of $(H^4/\dot\phi_0^2)(1/2k^3)$ due to the mixing with the new field ${s}$. The red, blue, green, orange and magenta curves are for $m = 0, 0.5H, H, 1.5$ and $2H$. The black dashed curve shows the result from the effective theory and the colored dashed lines are perturbation theory. }
\label{fig:funcmu}
\end{figure*}

Eq.~(\ref{eq:UVboundary}) is used to determine the initial conditions $\pi^{(1,2)}_k(\eta_0)$ , $s^{(1,2)}_k(\eta_0)$ and $\pi^{\prime (1,2)}_k(\eta_0)$, $s^{\prime(1,2)}_k(\eta_0)$ at a value of $\eta_0$ that is large in magnitude. The differential equations in (\ref{eq:diff1}) and (\ref{eq:diff2}) can then be solved numerically and used to determine the power spectrum for the curvature perturbation in this model. 

The correction to the power spectrum $\Delta{\cal P}_\zeta$  is defined by,  $\Delta{\cal P}_\zeta=  {\cal P}_\zeta-{\cal P}^{(0)}_\zeta$, where
\beq
{\cal P}_\zeta^{(0)}(k) = \frac{H^4}{\dot\phi_0^2} \frac{1}{2k^3} \ ,
\eeq
is the power spectrum of the curvature perturbation in usual slow roll single field inflation.
$\Delta{\cal P}_\zeta$ is shown in Fig.~\ref{fig:funcmu}.  In the region of $\mu \ll H$ $\Delta{\cal P}_\zeta$ goes like $\mu^2$ which agrees with the perturbative calculation~\cite{Chen:2009zp}.
In the region where $\mu$ is larger than about $10H$ the power spectrum ${\cal P}_\zeta$ grows as $\mu^{1/2}$ and can be approximated by,
\beq\label{eq:power}
{\cal P}_\zeta(k) = {\cal C} \left(\frac{\mu}{H} \right)^{1/2} {\cal P}_\zeta^{(0)}(k) \ ,
\eeq
where 
\beq
\label{normpower}
{\cal C} = \frac{16\pi}{\Gamma^2(-1/4)} \simeq 2.09.
\eeq
Corrections to eqs.~(\ref{eq:power}) and (\ref{normpower}) become negligible as $\mu \rightarrow \infty$. The power spectrum in the large $\mu$ limit was  calculated using the large $\mu$ effective field theory in~\cite{Gwyn:2012mw}. For completeness we briefly review that calculation in  Sec.~\ref{sec:effetivetheory}. 

As shown in Ref.~\cite{Chen:2009zp}, the perturbative result diverges in the limit of $m\rightarrow0$. From the red curve shown in Fig.~\ref{fig:funcmu} we can see that the curvature perturbation is well defined at $m=0$.  Perturbation theory can be very misleading at modest values of $m$ and values of $\mu$ not very much larger than unity. For example for $m=0.5H$ and $\mu=1.5H$ it gives a value for $\Delta {\cal P}_{\zeta}$ (in the units used for Fig.~\ref{fig:funcmu}) equal to  310 while our numerical result is 6.2.

\begin{figure*}[tp]
\centering
\includegraphics[width=3.5in]{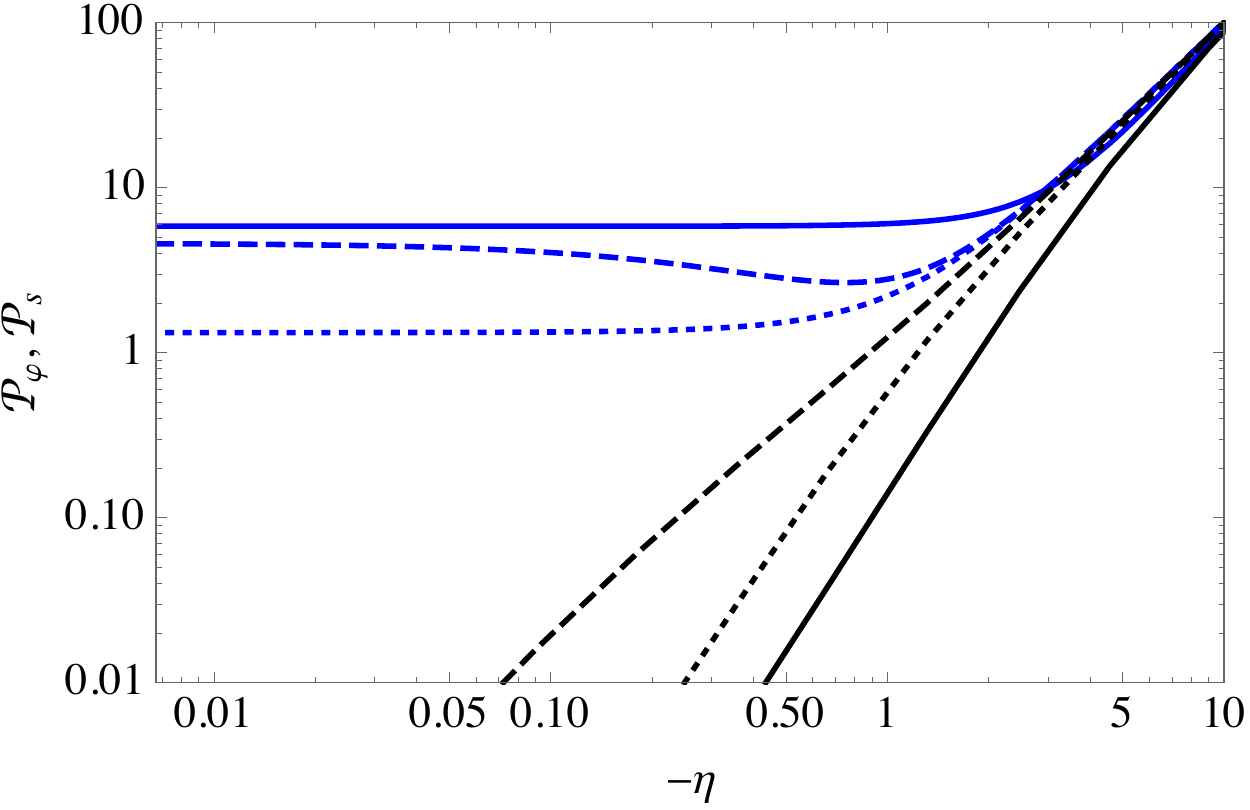}
\caption{Power spectrum of the $\pi$ and $s$ fields in the unit of $H^2/2k^3$. The solid, dotted, and dashed curves are for $(\mu/H, m/H) = (10,2), (1, 2)$ and $(1.2,0.9)$, respectively.}
\label{fig:s2}
\end{figure*}

For the curvature perturbations one calculates the power spectrum of the $\pi$ field as $-\eta \rightarrow 0$. However the power spectra for the fields can be calculated at any $\eta$. For $\mu/H > 1$ the power spectrum for the $s$ field ${\cal P}_s(k)$ falls off rapidly as $-\eta$ falls below unity. The numerical results of the power spectrum of the $s$ field ${\cal P}_s(k)$ in  units of $H^2/2k^3$ as a function of $\eta$ for a few values of $\mu$ and $m$ are shown in Fig.~\ref{fig:s2}. One can see that all the curves decrease  with $-\eta$ and become small as $-\eta$ falls below unity. 

In the usual single field inflation model ${\cal P}_\pi$ goes to unity in units of $H^2/2k^3$ as $-\eta \rightarrow 0$. In this model of quasi-single field inflation, as shown in Fig.~\ref{fig:s2} for the $\mu = 10H$, $m = 2H$ case the asymptotic value of ${\cal P}_\pi$ is  much larger than unity. This is due to the change in the dispersion relation of the $\pi$ field and can be understood  using the large $\mu$ effective theory. From Fig.~\ref{fig:s2} we see that the asymptotic value of ${\cal P}_\pi$ for the case $\mu = 1.2H, m = 0.9H$ is also much larger than 1. 

 \subsection{Qualitative analysis}
 \label{Qualitative analysis}
We can understand qualitatively the shape of the mode functions analytically.
In the region well outside the horizon, $-\eta \ll 1$, eqs.~(\ref{eq:diff1}) and (\ref{eq:diff2}) can be simplified to
\bea
&&-\pi''_k + \frac{2\pi'_k}{\eta} - \frac{\mu}{H} \left(\frac{3{s}_k}{\eta^2} - \frac{{s}'_k}{\eta}\right) = 0 \nn
&&-{s}''_k + \frac{2s'_k}{\eta} - \frac{m^2 s_k}{H^2 \eta^2} - \frac{\mu}{H} \frac{\pi'_k}{\eta} = 0
\eea
which is invariant under the transformation 
\beq
\pi_k \rightarrow \lambda^2\pi_k~,~~ {s}_k\rightarrow \lambda^2{s}_k~,~~ \eta\rightarrow \lambda\eta \ .
\eeq
Therefore, the general form of the solution can be written as
\beq
\pi_k = Q_k (-\eta)^{\alpha}~,~~ {s}_k = R_k (-\eta)^{\alpha} \ .
\eeq
Putting this back into the differential equations gives  equations for the power $\alpha$ and the coefficients $Q_k$ and $R_k$
\bea\label{eq:character}
(\alpha^2-3\alpha)Q_k + \frac{\mu}{H} (3 - \alpha) R_k &=& 0 \ ,\nn
\frac{\mu}{H} \alpha  Q_k + \left(\alpha^2-3\alpha + \frac{m^2}{H^2}\right) R_k &=& 0 \ .
\eea
To have nontrivial solutions for $Q_k$ and $R_k$ requires
\beq
\alpha (\alpha-3)\left[\alpha^2 - 3\alpha  +  \frac{m^2 + \mu^2}{H^2}\right] = 0 \ .
\eeq
There are four solutions to this equation
\beq
\label{alphas}
\alpha_1 = 0~,~~\alpha_2 = 3~,~~ \alpha_{\pm}  = \frac{3}{2}\pm \left(\frac{9}{4} - \frac{m^2 + \mu^2}{H^2}\right)^{1/2} \ .
\eeq
For the region of parameter space we focus on, $\alpha_{\pm}$ are complex, which can have observational consequences for the non-gaussianities~\cite{Arkani-Hamed:2015bza}.

\begin{figure*}[tp]
\centering
\includegraphics[width=3in]{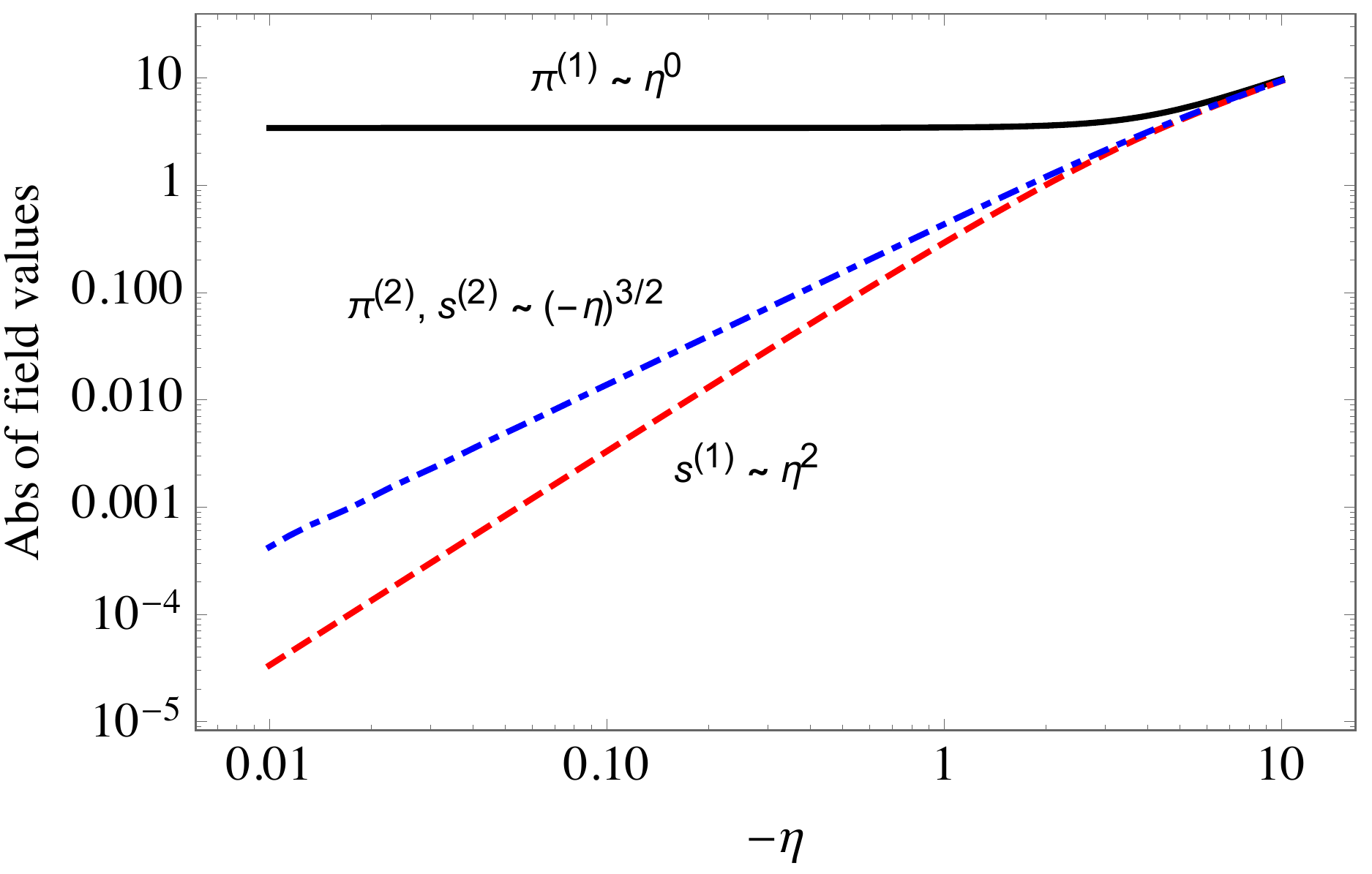}
\includegraphics[width=3in]{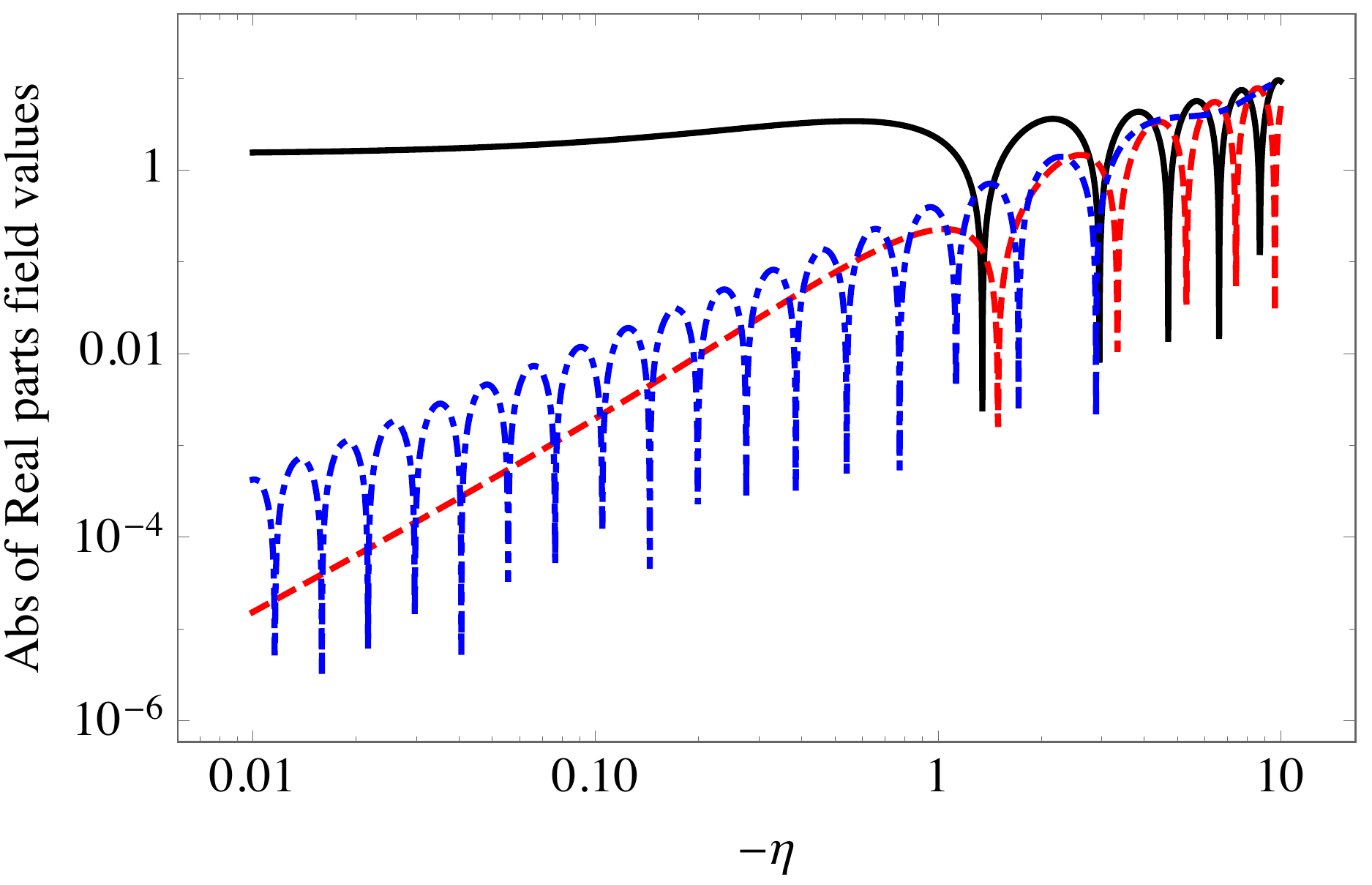}
\caption{Left: Absolute values of the field values with $m = 2H$ and $\mu = 10H$. The solid black and the dashed red curves are for the $\pi$ and ${s}$ mode with the index $\alpha = 0$. The dot-dashed blue curve illustrates the $\pi$ and $s$ modes whose dominant small $-\eta$ behavior comes from the index $\alpha = 3/2\pm (9/4- (m^2+\mu^2)/H^2)^{1/2}$. Right: Showing the absolute value of the real parts of each mode corresponding to the ones in the left panel.  }
\label{fig:solutionsIR}
\end{figure*}

For large values of $\mu/H$ the infrared behavior of the mode functions  $\pi_k^{(1,2)}$ and $s_k^{(1,2)}$  match directly onto the solutions in eq.~(\ref{alphas}).  This is shown in Fig.~\ref{fig:solutionsIR} using $m =2 H$ and $\mu = 10 H$. The $\alpha_1=0$ mode is constant outside the horizon. The $\alpha_2 = 3$ behavior  vanishes outside the horizon and can be thought of as a subdominant contribution to the massless mode. The $\alpha_{\pm}$ solutions correspond to the mode functions for a free scalar field with mass equal to $(m^2+\mu^2)^{1/2}$. They play an important role in the calculation of non-gaussianities. For $m= 2H$ and $\mu = 10H$ the behavior of this mode is shown by the blue dot-dashed curves in Fig.~\ref{fig:solutionsIR}. One can see that it oscillates logarithmically with frequency $(m^2+\mu^2)^{1/2}$, and decreases with a power of $3/2$ for small $-\eta$. To get the curves shown in Fig.~\ref{fig:solutionsIR} we solve the differential equations (\ref{eq:diff1}) and (\ref{eq:diff2}) with the initial conditions (\ref{eq:UVboundary}).  The $\pi^{(2)}$ mode shown in the left panel of Fig.~\ref{fig:solutionsIR} eventually goes to a constant as $-\eta$ gets smaller.  Similarly, the absolute value of the $s^{(1)}$ mode eventually goes like $(-\eta)^{3/2}$ for very small $-\eta$.  


In this paragraph we focus on the $\alpha_1 = 0$ solution. Putting $\alpha_1 = 0$ back Eq.~(\ref{eq:character}) we find that $R_k = 0$. Since there is no shift symmetry in the ${s}$ field  it should not contain the massless mode in the far infrared. We can get the leading behavior of the ${s}_k$ mode function outside the horizon by putting $\pi_k = Q _k$ back into the exact differential equation (\ref{eq:diff1}). This gives the first order inhomogeneous differential equation
\beq
- Q_k = \frac{\mu}{H} \left( \frac{3{s}_k}{\eta^2} - \frac{{s}_k'}{\eta}\right)
\eeq
with general solution 
\beq
\label{loveit}
{s}_k = -\frac{Q_k H\eta^2}{\mu} \ .
\eeq
This behavior is shown by the red dashed curves in Fig.~\ref{fig:solutionsIR}.

\subsection{The large $\mu/H$ region}

\begin{figure*}[tp]
\centering
\includegraphics[width=3.5in]{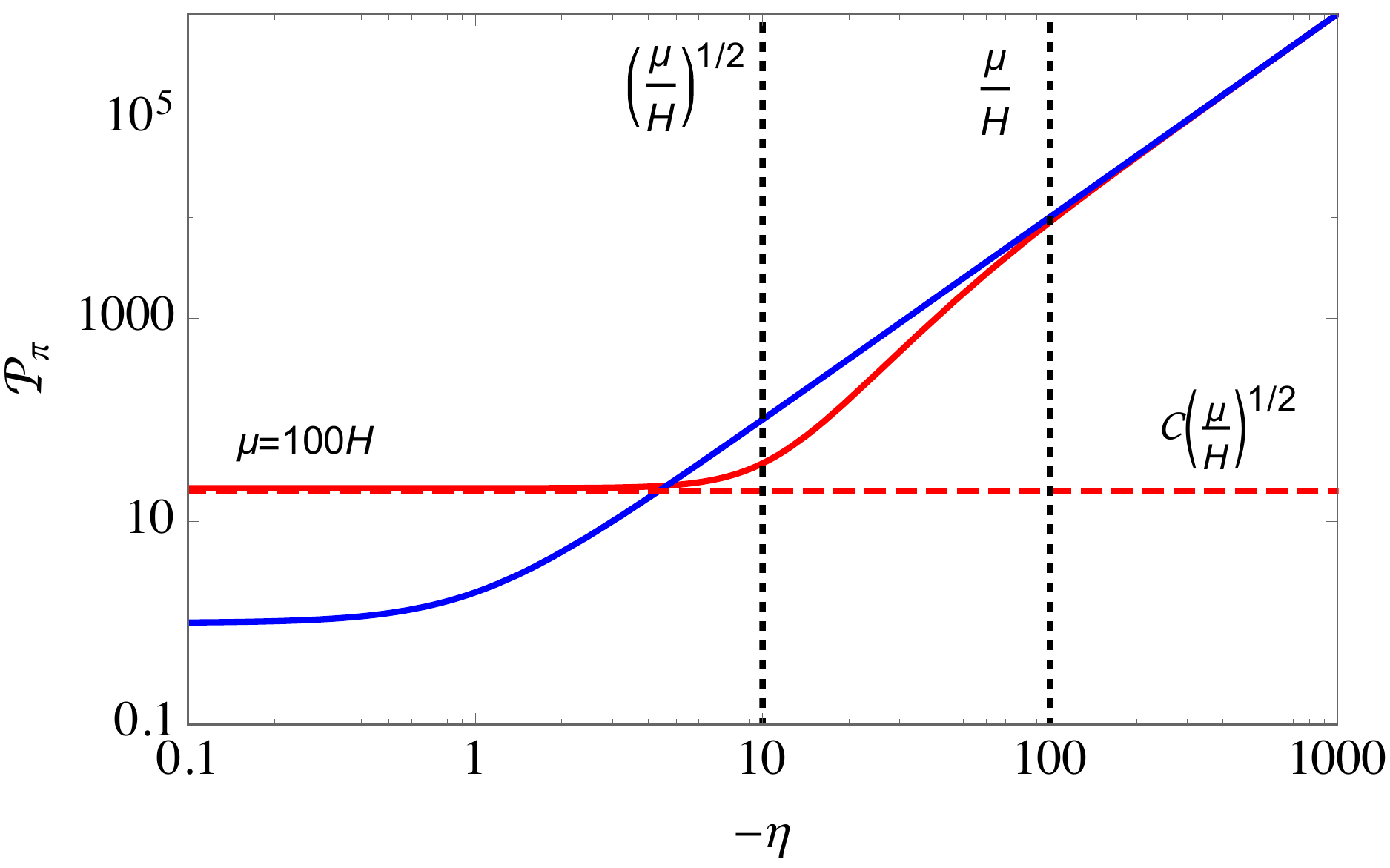}
\caption{Numerical result of ${\cal P}_\pi(k)$ as a function of $\eta$ in the unit of $ H^2/(2k^3)$ for $m=0$ and $\mu = 100 H$. For comparison in blue we show the result for standard single field inflation.}
\label{fig:plateau}
\end{figure*}

In this subsection we focus on some properties of the solutions for the mode functions that only apply for very large $\mu/H$. We find that the curvature perturbation goes  to a constant when $-\eta < (\mu/H)^{1/2}$ instead of the usual condition that it be outside the horizon, {\it i.e.}, $-\eta <1$.  This is illustrated in Fig.~\ref{fig:plateau}  which shows  the numerical results for the power spectrum of ${\cal P}_{\pi}$ as a function of $\eta$.

Examining  eq.~(\ref{eq:diff2}), in the region $-\eta < (\mu/H)^{1/2}$ it is clear that the last term on the left hand side is the largest. Neglecting the other terms the solution in this region satisfies
\beq
\pi' = 0  \ ,
\eeq
which implies that $\pi_k$ is constant and $s_k$ is proportional to $\eta^2$, as in eq.~(\ref{loveit}).

In the region $(\mu/H)^{1/2}< -\eta < \mu/H$ one can show  that the differential equations for the mode functions are solved approximately by 
\beq
\pi_k \propto  (-\eta)^{3/2} \exp \left[  \frac{iH \eta^2}{2\mu} \right] \ ,\;\; s_k \propto  (-\eta)^{3/2} \exp \left[  \frac{iH \eta^2}{2\mu} \right] \ .
\eeq
The physical wavevector of a mode with comoving wavevector $k$ is 
\beq
q = k a^{-1} = - k H \tau \ .
\eeq
Therefore the change of the phase of these solutions  within a small time period $\Delta\eta$ can be written as
\beq
{\Delta}{\rm phase}  = \frac{\eta_0 \Delta \eta H}{\mu} = -\frac{q^2}{\mu } \Delta t \ ,
\eeq
where $\Delta t = a \Delta\tau $ has been used. This agrees with the dispersion relation in flat space given in eq.~(\ref{eq:flatdispersion}) for the massless mode. 
From Fig.~\ref{fig:plateau}, one can see that it is in this region the solution for $\mu \gg H$ starts to deviate from the standard slow roll solution, which corresponds to $\mu=0$ in the model we are studying. This is because in this region the solutions in de-Sitter space should resemble those in flat space and the light  mode has a flat space dispersion relation $E_q = q^2/\mu$ which is quite different from a single massless field with dispersion relation $E_q = q$. 

Putting the solution we have found back into the differential equations (\ref{eq:diff1}) and (\ref{eq:diff2}), one can see that the terms 
\beq
- \pi''_k + \frac{2\pi'_k}{\eta} ~~{\rm and}~~ - s''_k + \frac{2 s'_k}{\eta} \ 
\eeq
are suppressed, which means that the terms
\beq
(\partial_\tau\pi)^2 ~~{\rm and}~~  (\partial_\tau s)^2  \
\eeq
in the Lagrange density (\ref{eq:Lagrange}) can be neglected. After neglecting these two terms, there are no terms in  (\ref{eq:Lagrange}) that contain time derivatives of $s$.  This indicates that $s$ has become a Lagrange multiplier and can be replaced in the Lagrange density using its classical equation of motion to express it in terms of $\pi$. This amounts to summing the tree graphs that contain virtual $s$ propogators and is the origin of the  effective theory approach developed in Refs.~\cite{Baumann:2011su} and \cite{Gwyn:2012mw} for the behavior of $\pi$ in this region. We will briefly review the basic setup for this effective field theory and use it to calculate the two- and three-point functions of the curvature perturbation in the large $\mu$ limit in Sec.~\ref{sec:effetivetheory}.

\section{Impact on observables}
\label{sec:observable}

The dimensionless power spectrum is defined as~\cite{Ade:2015xua} 
\beq\label{eq:DeltaS}
\Delta^2_\zeta(k) = \frac{k^3}{2\pi^2} {\cal P}_\zeta(k) = \frac{H^4}{(2\pi)^2\dot\phi_0^2}  f(\mu/H, m/H) = 2.12\times 10^{-9} \ ,
\eeq
where $f$ is a function of the $\mu$ and $m$. $f-1$  is shown in Fig.~\ref{fig:funcmu} as a function of $\mu/H$ for fixed values of $m$. Throughout this section we neglect the impact of the time dependence of ${\dot \phi}_0$ on the value of $S_0$ since, as was discussed in section~\ref{The Model},  it is  suppressed by a power of $\Lambda$.

In terms of the slow-roll parameter
\beq
\epsilon = \frac{\dot\phi_0^2}{2 H^2 M_{\rm pl}^2} \ 
\eeq
$\Delta^2_\zeta(k)$ can be written as
\beq
\Delta^2_\zeta(k) = \frac{H^2}{8\pi^2 \epsilon M_{\rm pl}^2} f\left(\frac{\sqrt{8\epsilon} M_{\rm pl}}{\Lambda}, \frac{m}{H}\right) \ .
\eeq
The tilt of the power spectrum is defined as
\beq
n_S - 1 \equiv \frac{d\log\Delta_\zeta^2}{d\log k} \ ,
\eeq
and can be written as
\beq
n_S - 1 = \frac{d\log\Delta_\zeta^2}{d\log k} = \frac{d\log\Delta_\zeta^2}{d N}\times \frac{d N}{d\log k} \ ,
\eeq
where $N$ is the number of e-folds between when the modes of interest exit the horizon and inflation ends. From Eq.~(\ref{eq:DeltaS}) we have
\bea\label{eq:dDeltadN}
\frac{d\log \Delta_\zeta^2}{d N } &=& 2\frac{d\log H}{d N} - \frac{d\log\epsilon}{dN} + \left( \frac{\partial\log f}{\partial\log\hat \mu} \frac{d\log\hat\mu}{d\log\epsilon}\frac{d\log\epsilon}{dN} + \frac{\partial \log f}{\partial \log\hat m} \frac{d\log\hat m}{d\log H} \frac{d\log H}{d N} \right) \nn 
&=& -4\epsilon+2 \eta + (\epsilon - \eta) \frac{\partial \log f}{\partial \log\hat\mu} + \epsilon \frac{\partial \log f}{\partial\log\hat m}
\eea
where the standard results of slow-roll inflation have been used~\cite{Baumann:2009ds}, and $\eta$ is the other slow-roll parameter defined as $- \ddot\phi_0/(H\dot\phi_0)$. $\hat \mu$ and $\hat m$  are defined as
\beq
\hat \mu \equiv \frac{\mu}{H} \ ,\;\; \hat m \equiv \frac{m}{H} \ .
\eeq
Up to leading order in the slow-roll parameters we have that,
\beq
\frac{d \log k}{d N} = 1 
\eeq
Therefore at  leading order in slow roll parameters
\beq
n_S - 1 = -4\epsilon+2 \eta + (\epsilon - \eta) \frac{\partial \log f}{\partial \log\tilde\mu} + \epsilon \frac{\partial \log f}{\partial\log\tilde m} \ .
\eeq 

Another important observable is the tensor-scalar ratio. Since the gravitational wave production is only related to the structure of the de-Sitter metric, the dimensionless tensor spectrum can  still be written as
\beq
\Delta_t^2 = \frac{2}{\pi^2}\frac{H^2}{M_{\rm pl}^2} \ .
\eeq
Then the tensor-scalar ratio can be written
\beq
r = \frac{\Delta_t^2(k)}{\Delta^2_\zeta(k)} = 16 \epsilon \times f^{-1}(\hat\mu, \hat m)  \ .
\eeq

\subsection{$\phi^2$ inflation}

\begin{figure*}[tp]
\centering
\includegraphics[width=4in]{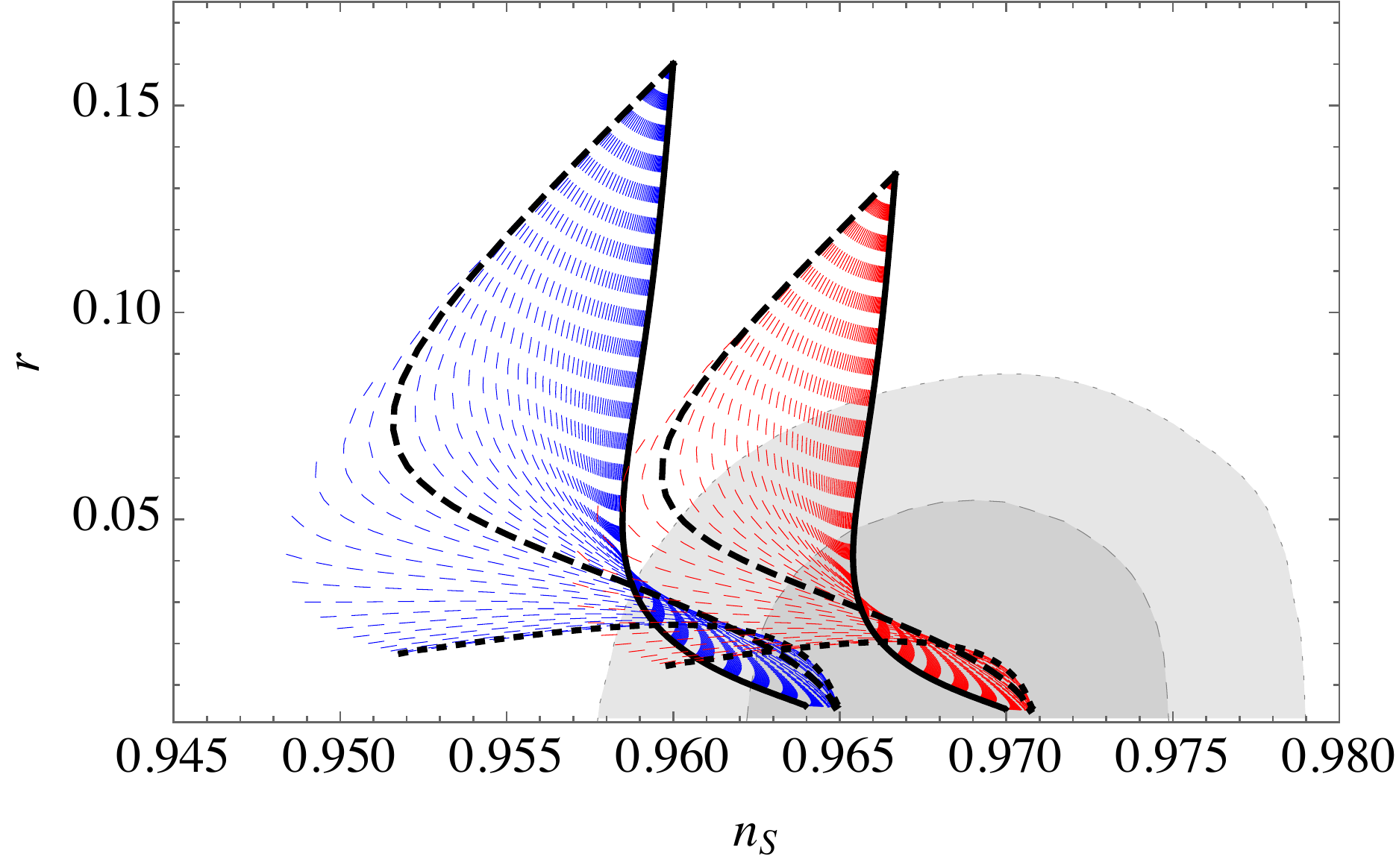}
\caption{Impact on the scalar spectrum index $n_S$ and the tensor-to-scalar ratio $r$ for the $\phi^2$ inflation model with $\mu$ from 0 to 100$H$ and $m$ from 0 to 6$H$, and  $(\mu^2 + m^2)^{1/2} > 3H/2$. The blue and red regions are for $N_{\rm cmb} = 50$ and 60 respectively. The dotted, dashed and solid curves are for $m$ fixed to be $0,3H/2$ and $6 H$ respectively.  The gray regions are the one-sigma and two-sigma constraints from the combination of the Planck data and the BICEP2/Keck data~\cite{Array:2015xqh}.}
\label{fig:nsr}
\end{figure*}

\begin{figure*}[tp]
\centering
\includegraphics[width=4in]{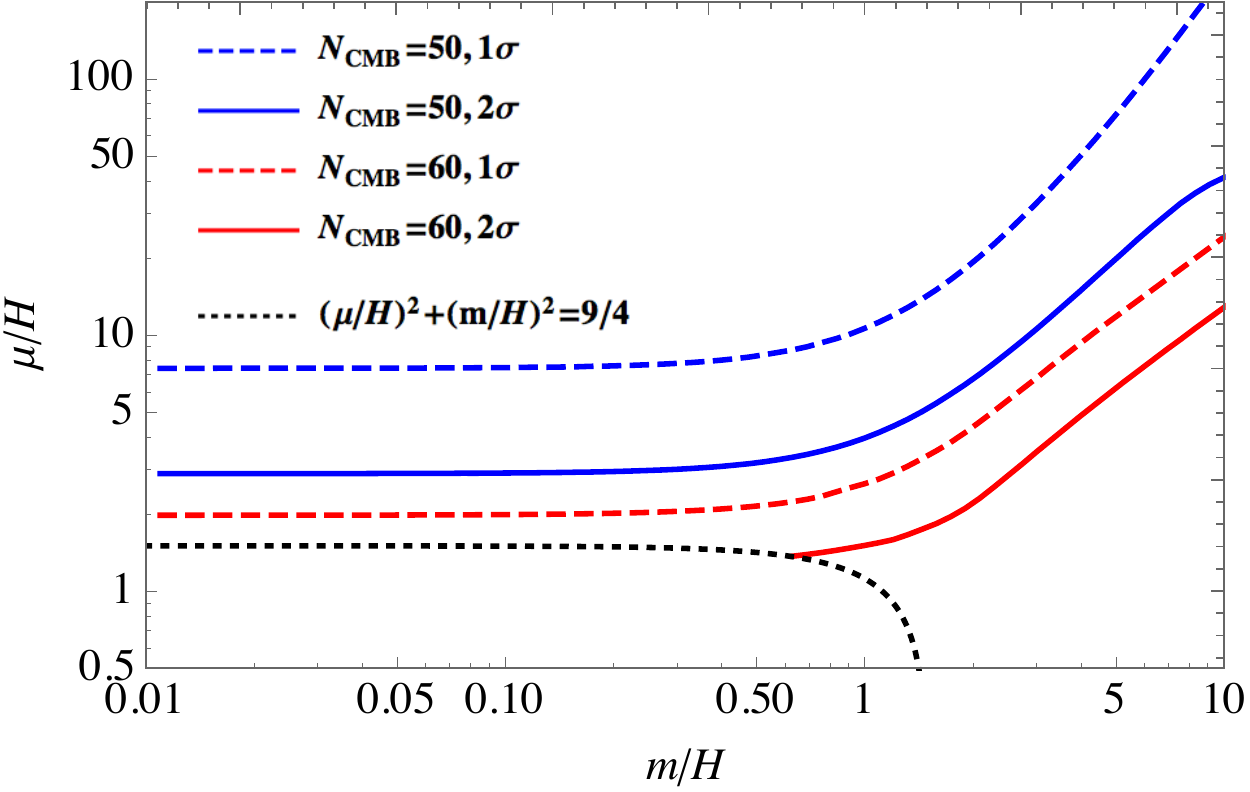}
\caption{Constraints on the $m-\mu$ parameter space from the combination of the Planck data and the BICEP/Keck data~\cite{Array:2015xqh}, where the blue curves are for $N_{\rm CMB}=50$ and the red curves $N_{\rm CMB}=60$. The regions above the curves are allowed.}
\label{fig:mmu}
\end{figure*}

Here we use the model where the inflaton potential $V_{\phi}=m_{\phi}^2\phi^2/2$  as an example to discuss the effect of large $\mu$ on the observables. In this simple model, we have 
\beq
\phi_{\rm cmb} = 2\sqrt{N_{\rm cmb}} M_{\rm pl} \simeq 15 M_{\rm pl} \ .
\eeq
and
\beq
\epsilon = 2\left(\frac{M_{\rm pl}}{\phi_{\rm cmb}}\right)^2 \simeq \frac{1}{2 N_{\rm cmb}} \ , ~~{\eta} \simeq 0 \ ,
\eeq 
where $N_{\rm cmb}$ is the number of e-folds between when CMB scale leaves  the horizon and when slow roll inflation ends.

The $n_S,r$ plot for this model is  shown in Fig.~\ref{fig:nsr}. The dotted regions are for $\mu$ from 0 to $100 H$ and $m$ from $0$ to $6H$ with $(\mu/H)^2+(m/H)^2 > 9/4$.  On these curves as  $\mu$ increases $r$ decreases, so the uppermost point of the  curves corresponds to standard slow roll inflation. 
The constraints on the $m-\mu$ parameter space for $N_{\rm CMB}=50$ and 60 are also shown in Fig.~\ref{fig:mmu} where the regions below the curves are excluded. Clearly larger values of $\mu$ improve the agreement of the model's predictions with the measured value of $n_S$ and the bound on $r$.
%
%


\section{Non-Gaussianities}
\label{sec:nonGaussian}

In this section we calculate the dependence of the inflaton three-point function as a function of $\mu$ and $m$.  The small $\mu$ behavior of the bispectrum was first studied in \cite{Chen:2009zp}.  The effective field theory for large $\mu$ was used to compute the contribution from the $\partial \pi \partial \pi s$ interaction to the bispectrum ~\cite{Baumann:2011su,Gwyn:2012mw}. Here we use the numerical mode functions to extend the analysis to other values of $\mu$.  

The curvature perturbation bispectrum $B_\zeta({\bf k}_1, {\bf k}_2,{\bf k}_3)$ is defined by
\begin{equation}
\label{bizeta}
	\langle \zeta({\bf x}_1,0)\zeta({\bf x}_2,0)\zeta({\bf x}_3,0)\rangle= \int\frac{d^3k_1}{(2\pi)^3}\frac{d^3k_2}{(2\pi)^3}\frac{d^3k_3}{(2\pi)^3}e^{i ({\bf k}_1\cdot{\bf x}_1 +{\bf k}_2 \cdot{\bf x}_2+{\bf k}_3\cdot{\bf x}_3)}B_\zeta({\bf k}_1,{\bf k}_2,{\bf k}_3)(2\pi)^3\delta^3({\bf k}_1+{\bf k}_2 +{\bf k}_3)
\end{equation}
and we can define $B_\pi({\bf k}_1,{\bf k}_2,{\bf k}_3)$ analogously.
They can be computed using the in-in formalism \cite{Weinberg:2005vy} using the interaction Lagrangian in eq.~(\ref{intlagrange}).  



In this section we  focus mostly on the $O(V_S''')$ term (where ${V_S'''} \equiv V_S'''(S_0)$) which, for $V_S''' \sim {\cal O}(H)$, typically dominates over the contribution from the $\partial \pi \partial \pi s$ term. 
We express the $O(V_S''')$ contribution to the bispectrum in terms of the mode functions discussed earlier.  Evaluating the correlator in the far future $\tau=0$, we find
\begin{align}
\label{analytic three point}
B_\pi({\bf k}_1,{\bf k}_2,{\bf k}_3) &= -2V_S'''H^{-4}\textrm{Im}\left[\int_{-\infty}^{0}\frac{d\tau}{\tau^{4}}\prod_{i=1}^{3}\left(\pi_{k_i}^{(1)*}(0)s_{k_i}^{(1)}(k_{i}\tau) + \pi_{k_i}^{(2)*}(0)s_{k_i}^{(2)}(k_{i}\tau) \right)\right].
\end{align}
Equation (\ref{analytic three point}) is true for all values of $k_{i}$, however we are mostly interested in its behavior in the so-called equilateral and squeezed limits.  In the equilateral limit, the external momenta all have equal magnitude $k_{i} \equiv k$.  In this case, the integral's dependence on $k$ can be factored out of the integral by rescaling the integration variable from $\tau$ to $\eta=k\tau$:\footnote{By, $B_\pi^{{\rm equil}}(k)$, we mean $B_\pi$ evaluated in the equilateral configuration where the three wavevectors have the same magnitude $k$.} 
\begin{align}
\label{equilateral analytic}
B_\pi^{\text{equil}}(k) &= -2V_S'''H^{-4}k^3\textrm{Im}\left[\int_{-\infty}^{0}\frac{d\eta}{\eta^{4}}\left(\pi_{k}^{(1)*}(0)s_{k}^{(1)}(\eta) + \pi_k^{(2)*}(0)s_k^{(2)}(\eta) \right)^{3}\right]
\end{align}

We can compute this integral numerically using the numeric mode functions, but there are a couple of subtleties in its evaluation that need to be addressed. The integrand in (\ref{equilateral analytic}) is highly oscillatory at large $\tau$.  For $m/H$ and $\mu/H$ values of order one or larger, the magnitude of these oscillations does not decay quickly and it becomes difficult to perform the numerical integrations by brute force.  We can alleviate this problem by Wick rotating the integral, thereby transforming the rapid oscillations into exponential decay.

Before Wick rotating it is convenient to factor out the oscillatory behavior from the mode functions.  The large $\tau$ limit given in eq.~(\ref{large tau limit}) suggests that we should extract the oscillatory behavior by factorizing the mode functions as $\pi_k^{(i)}(\eta) = A_k^{(i)}(\eta)e^{-i\eta}$ and $s_k^{(i)}(\eta) = B_k^{(i)}(\eta)e^{-i\eta}$.  Plugging this factorization into $B_\pi^{\text{equil}}(k)$ gives
\begin{align}
\label{lambda equil}
B_\pi^{\text{equil}}(k) &=  -2V_S'''H^{-4}k^3\textrm{Im}\left[ \int_{-\infty}^{0}\frac{d\eta}{\eta^{4}}e^{-3i\eta}\left(\pi_k^{(1)*}(0)B_k^{(1)}(\eta) + \pi_k^{(2)*}(0)B_k^{(2)}(\eta) \right)^{3}\right]\cr
 &= -2V_S'''H^{-4}k^3\textrm{Re}\left[\int_{-\infty}^{0}\frac{dx}{x^{4}}e^{3x}\left(\pi_k^{(1)*}(0)B_k^{(1)}(i x) + \pi_k^{(2)*}(0)B_k^{(2)}(i x) \right)^{3}\right].
\end{align}
In the second line we used Cauchy's theorem to rotate the region of integration from the real to the imaginary axis and changed the integration variable from $\eta$ to $x =-i\eta$.

The numerical solutions found previously for $A_k^{(i)}(\eta)$ and $B_k^{(i)}(\eta)$ are functions of the real variable $\eta$ and cannot be integrated along the imaginary axis.  However, we can analytically continue them to the imaginary axis by Wick rotating the original mode equations (\ref{eq:diff1}) and (\ref{eq:diff2}) (see \cite{Assassi:2013gxa}).  After factoring out the oscillatory behavior and changing variables to $x=-i \eta$, we find that the analytically continued  functions $A_k^{(i)}$ and $B_k^{(i)}$ obey 
\begin{align}
\label{factorized wick}
&x^2A_k''(ix)+(2x^2-2x)A_k'(ix)-2x A_k(i x)-\frac{\mu}{H} x B_k'(i x)+(3- x)\frac{\mu}{H} B_k(i x)=0\\
&x^2 B_k''(i x)+(2x^2-2x)B_k'(ix)+\left(\frac{m^2}{H^{2}}-2x\right) B_k(i x)+\frac{\mu}{H} x A_k'(ix)+\frac{\mu}{H} x A_k(ix)=0
\end{align}
where a prime denotes a derivative with respect to $x$ and we have dropped the superscripts for simplicity.  The solutions should asymptote at large $-x$ to\footnote{If we hadn't first extracted the oscillatory factor, an exponentially suppressed factor would have appeared in (\ref{factorized wick initial}) that would have made the boundary conditions too small to solve (\ref{factorized wick}) numerically.}
\begin{align}
\label{factorized wick initial}
&A_k^{(1)}(ix) = \frac{H}{2k^{3/2}}(-ix)^{1+i\mu/2H}\ \ \ A_k^{(2)}(ix) = \frac{H}{2k^{3/2}}(-ix)^{1-i\mu/2H}\cr
&B_k^{(1)}(ix) =\frac{iH}{2k^{3/2}}(-ix)^{1+i\mu/2H}\ \ \ B_k^{(2)}(ix) =\frac{-iH}{2k^{3/2}}(-ix)^{1-i\mu/2H}.
\end{align}
These solutions and their derivatives with respect to $x$ give the initial conditions for numerical integration of the differential equations for $A_k$ and $B_k$.  Note that $A_k^{(i)}$ and $B_k^{(i)}$ contain an overall factor of $k^{-3/2}$.  Moreover, $\pi_k^{(i)}$ and $s_k^{(i)}$ have the same $k$-dependent normalization.  This implies that $B_\zeta^\text{equil}(k)/\mathcal{P}_\zeta^\text{equil}(k)^2$ is $k$-independent.

 \begin{figure*}[]
\includegraphics[width=4in]{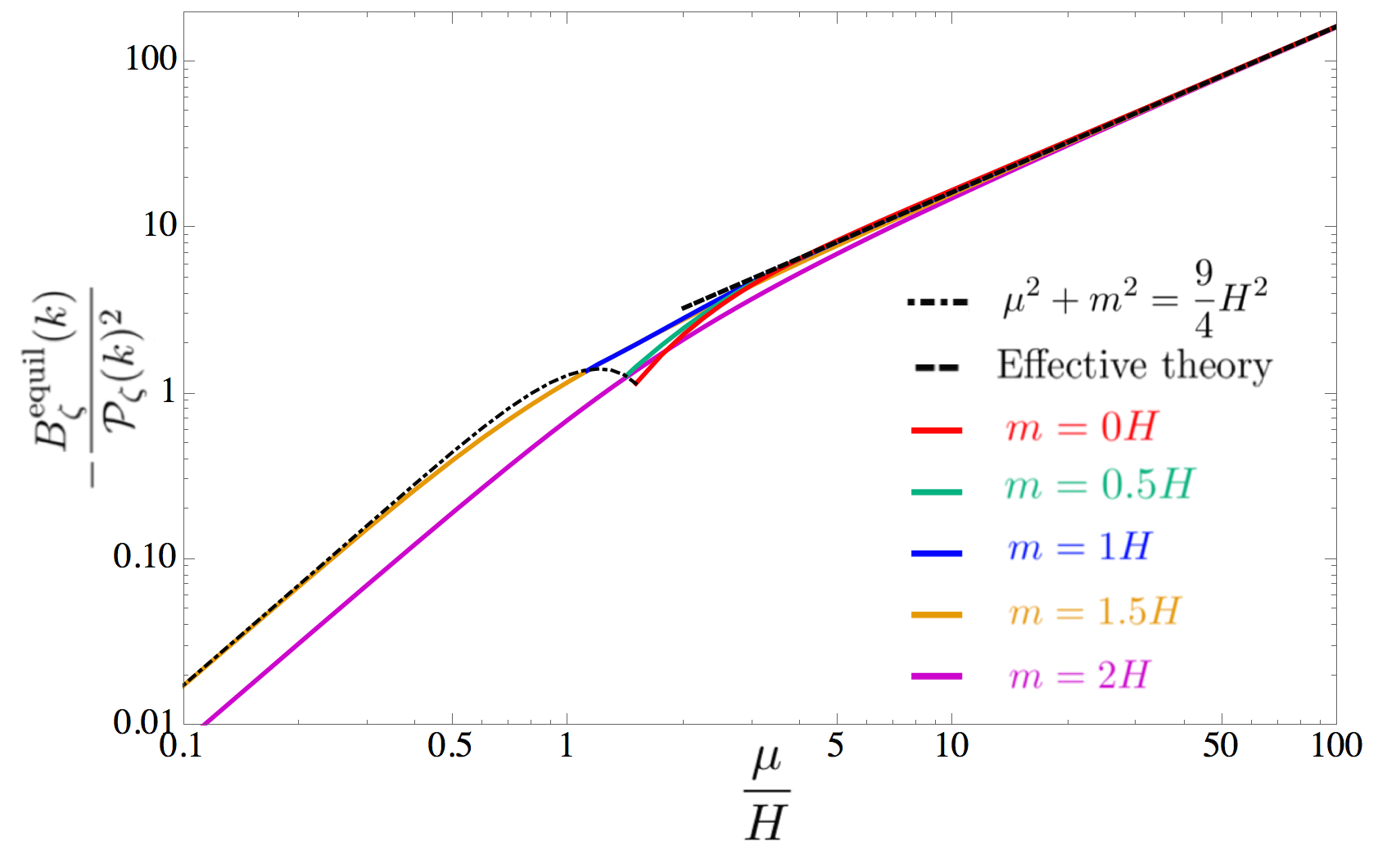}
\caption{The scaled equilateral three-point function due to the $s\partial\pi\partial\pi$ interaction, $B_\zeta^\text{equil}(k)/\cal{P}$$_\zeta( k )^2$ as a function of $\mu$.  Several values of $m$ are plotted: $m=0$, $0.5H$, $H$, $1.5H$, and $2H$, and there is also a black dashed line representing the result computed in the large $\mu$ effective theory.}
\label{fig:3ptfunction1}
 \includegraphics[width=4in]{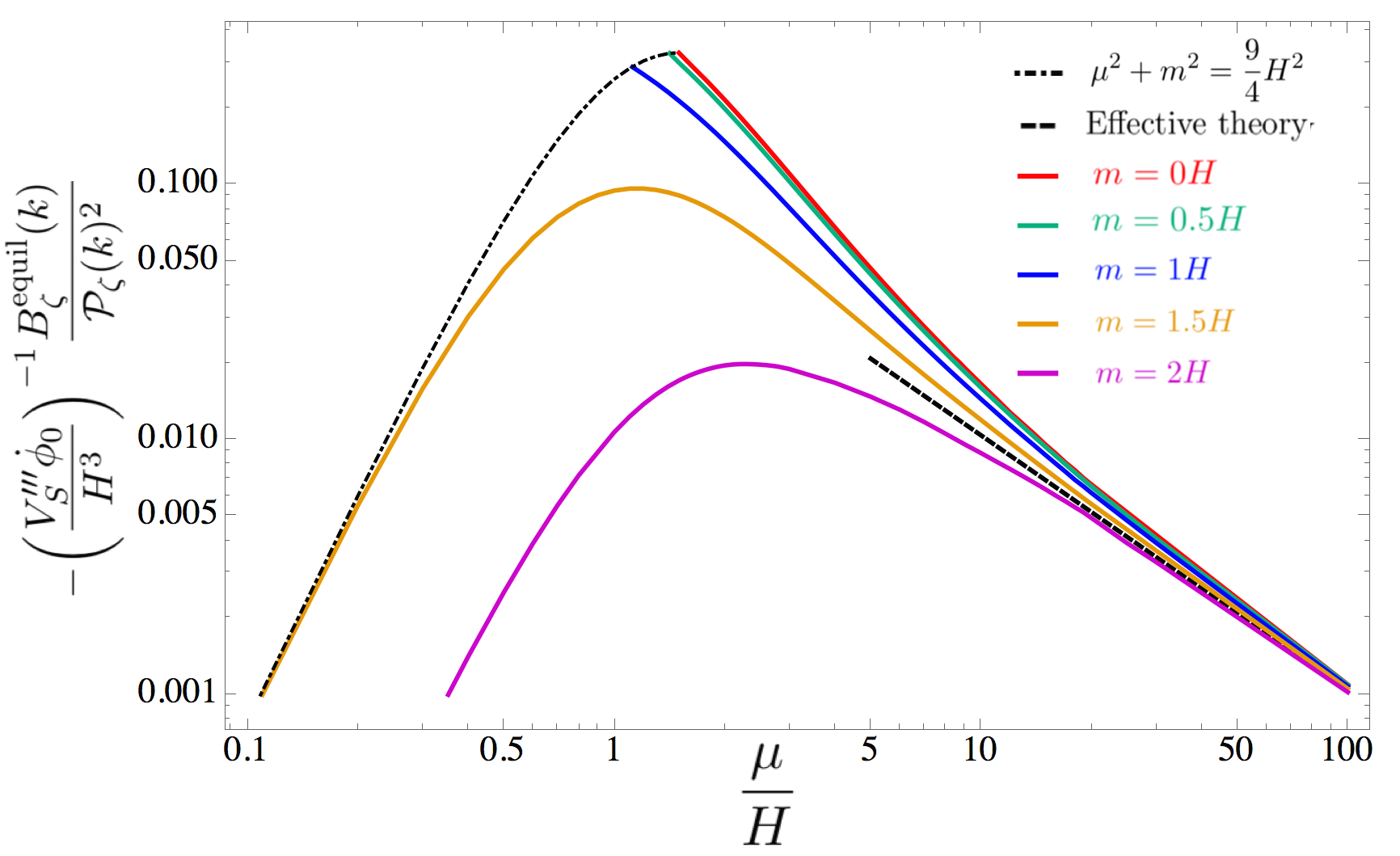}
\caption{The scaled equilateral three-point function due to the $s^3$ interaction, $B_\zeta^\text{equil}(k)/\cal{P}$$_\zeta( k )^2$ as a function of $\mu$.  Several values of $m$ are plotted: $m=0$, $0.5H$, $H$, $1.5H$, and $2H$, and there is also a black dashed line representing the result computed in the large $\mu$ effective theory.}
\label{fig:3ptfunction2}
\end{figure*}

In figures~\ref{fig:3ptfunction1} and~\ref{fig:3ptfunction2}, we plot the contributions to the scaled equilateral three-point functions $B^{\rm equil}_\zeta(k)/({\cal P}_{\zeta}(k))^2$ due to the $\partial\pi\partial\pi s$ and  $s^3$ interaction terms respectively\footnote{For brevity we have not described in any detail the calculation of the contribution due to the $s\partial\pi\partial\pi$ term in this section.}.  Moreover, we have superimposed a dotted line which corresponds to the prediction of the effective field theory appropriate for large $\mu$ (which will be discussed in detail in section \ref{sec:effetivetheory}).  Of course, the numerical results converge to the effective field theory results in the large $\mu$ limit.  However, the effective field theory is only a good approximation of these non-gaussianities for $\mu\gtrsim10H$.  This further suggests that there is a substantial portion of the parameter space in $\mu$ that is described neither by the large $\mu$ effective theory description nor the small $\mu$ perturbative description.

\begin{figure*}[]
\includegraphics[width=4in]{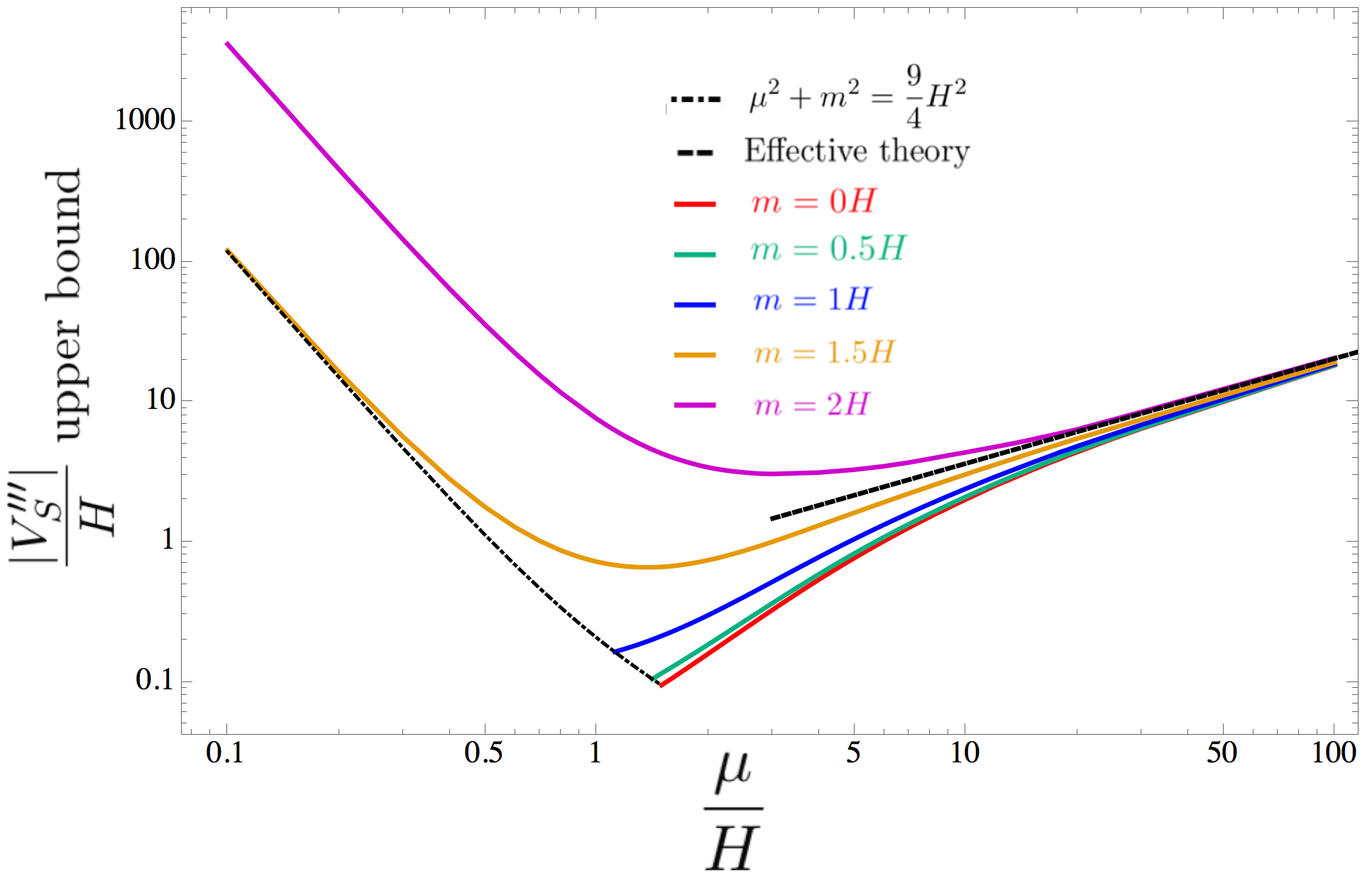}
\caption{Upper bounds on $|V_S'''|$ as a function of $\mu$.  These bounds are imposed by experimental bounds on $f_\text{NL}^\text{equil}$.  Bounds are plotted for $m=0$, $0.5H$, $H$, $1.5H$, and $2H$.  There is also a bound computed from the large $\mu$ effective theory, shown in the figure as a black dashed line. }
\label{fig:Vbounds}
\end{figure*}

The Planck collaboration has derived constraints on the magnitude of the bispectrum of the curvature perturbations using various models/templates for its dependence on the wavevectors \cite{Ade:2015ava}. These are usually expressed in terms of the quantity $f_\text{NL}$.  Although the model we are discussing is different from the equilateral model/template used to derive the constraint $f_\text{NL}^\text{equil}=4\pm43$ by the Planck collaboration in Ref.~\cite{Ade:2015ava}, we use this constraint to estimate a bound on $V_S'''$. Furthermore we estimate  $f_{\rm NL}^{\rm equil}$  using just the equilateral configuration where the three wavevectors have the same magnitude taking,
\beq 
 f_{\rm NL}^{\rm equil} \simeq \frac{5}{18}\times \frac{B^{\rm equil}_{\zeta}(k)}{({\cal P}_{\zeta}(k))^2}.
 \eeq
 To determine upper bounds for $V_S'''$ we assume that each interaction $s^3$ and $s\partial\pi\partial\pi$ is separately constrained by $f_\text{NL}^\text{equil}$ and thus ignore any possible tuning between the two terms that may make these bounds weaker.  Figure~\ref{fig:Vbounds} shows the $2\sigma$ upper bounds for a variety of  $s$ masses, as well as the upper bound predicted in the large $\mu$ effective theory.

\begin{figure*}[]
\includegraphics[width=4in]{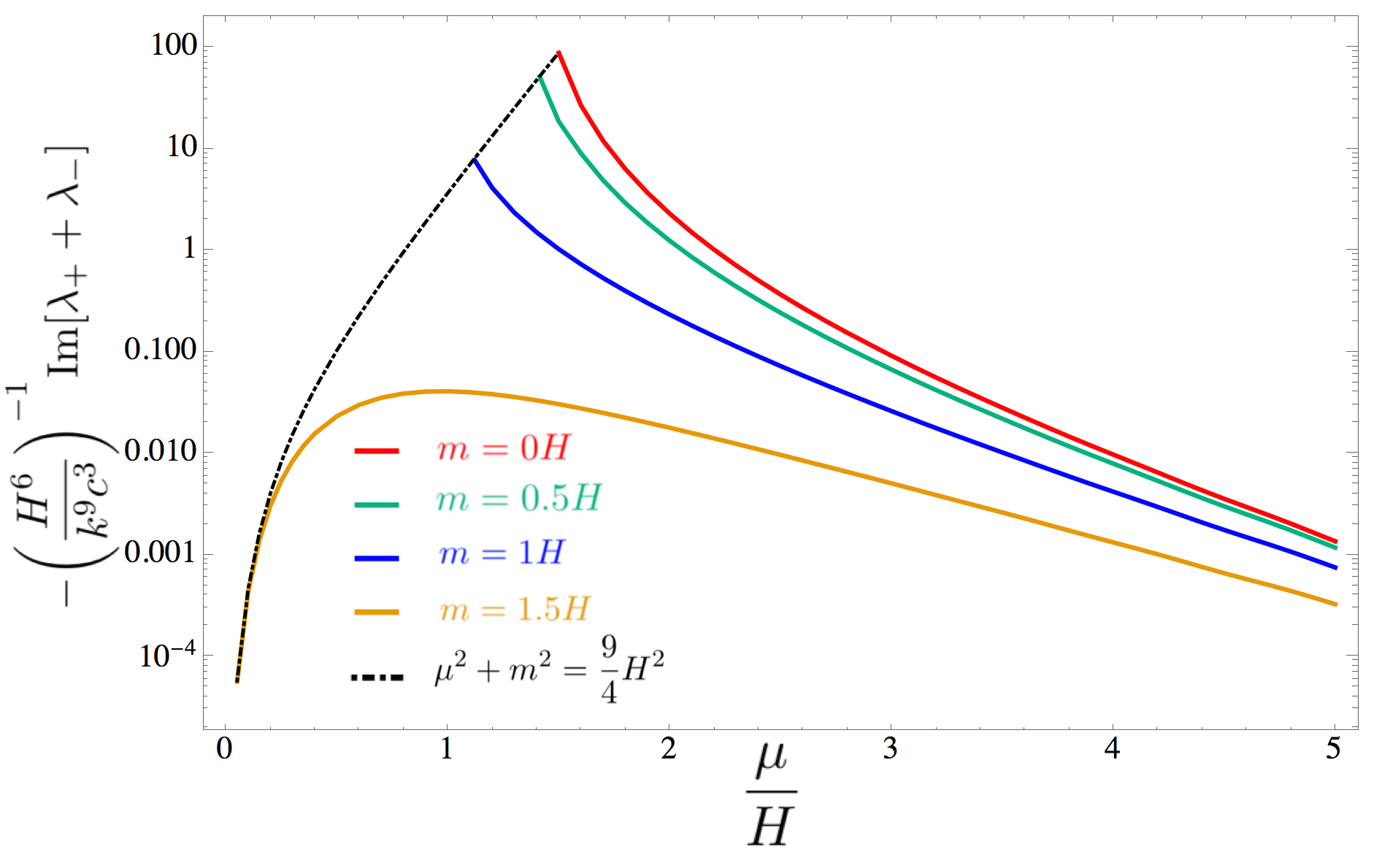}
\caption{The coefficients of the cosine term in equation (\ref{squeezed lambda im limit}) for $m=0$, $0.5H$, $H$, and $1.5H$.}
\label{fig:squeezedcos}
\end{figure*}
\begin{figure*}[]
 \includegraphics[width=4in]{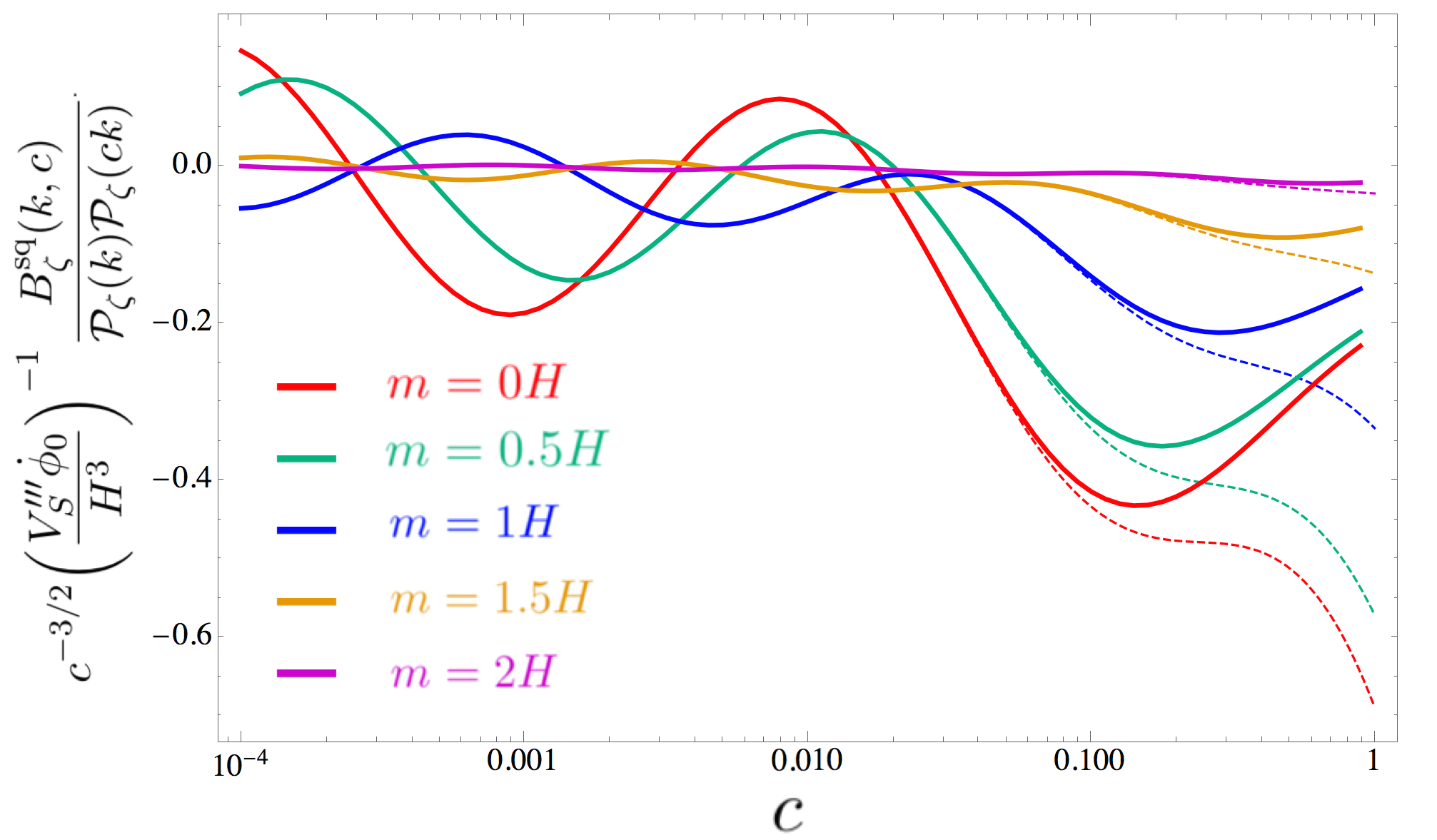}
\caption{In the squeezed limit, the three-point function logarithmically oscillates as a function of c.  This behavior is illustrated for $\mu=2$ and $m=0$, $0.5H$, $H$, $1.5H$, and $2H$.  The solid lines show the exact behavior as a function of $c$ ({\it i.e.} using equation (\ref{squeezed three point})) whereas the dotted lines show the approximate behavior to quadratic order in $c$ ({\it i.e.} using equation (\ref{squeezed integrand leading c behavior})).  }
\label{fig:squeezedoscillations}
\end{figure*}

The squeezed limit of (\ref{analytic three point}) occurs when $k_{1}\approx k_{2} \equiv k \gg k_{3}$. In this limit, define the ratio $c\equiv k_3/k$, where $c \ll1$,  and introduce the notation $B^{\rm sq}_{\pi}(k,c)$ for $B_{\pi}$.  We  again rescale the integration variable to $\eta=k\tau$ to find
\begin{align}
\label{squeezed three point}
B^{\rm sq}_\pi(k,c) &= -2V_S'''H^{-4}k^3\textrm{Im}\bigg[\int_{-\infty}^{0}\frac{d\eta}{\eta^{4}}\left(\pi_k^{(1)*}(0)s_k^{(1)}(\eta) + \pi_k^{(2)*}(0)s_k^{(2)}(\eta) \right)^{2}\cr
&\ \ \ \ \ \ \ \ \ \ \ \ \ \ \ \ \ \ \ \ \ \ \ \ \ \ \ \ \ \ \ \ \times\left(\pi_{c k}^{(1)*}(0)s_{ck}^{(1)}(c\eta) + \pi_{ck}^{(2)*}(0)s_{ck}^{(2)}(c\eta)\right)\bigg]\nonumber\\\cr\cr
& = -2V_S'''H^{-4}k^3c^{-3}\textrm{Im}\bigg[\int_{-\infty}^{0}\frac{d\eta}{\eta^{4}}\left(\pi_k^{(1)*}(0)s_k^{(1)}(\eta) + \pi_k^{(2)*}(0)s_k^{(2)}(\eta) \right)^{2}\cr\
&\ \ \ \ \ \ \ \ \ \ \ \ \ \ \ \ \ \ \ \ \ \ \ \ \ \ \ \ \ \ \ \ \times\left(\pi_{ k}^{(1)*}(0)s_{k}^{(1)}(c\eta) + \pi_{k}^{(2)*}(0)s_{k}^{(2)}(c\eta)\right)\bigg]
\end{align}
We can analyze the leading behavior of (\ref{squeezed three point}) in $c$ by replacing $s_k^{(i)}(c\eta)$ with the first few terms of its power series expansion (see section~\ref{Qualitative analysis}) $b_-^{(i)}(-c\eta)^{\alpha_-}+b_+^{(i)}(-c\eta)^{\alpha_+}+b_2^{(i)}(-c\eta)^2$:
\begin{align}
\label{squeezed integrand leading c behavior}
B_\pi^\text{sq}(k,c) &= -2V_S'''H^{-4}k^3c^{-3}\textrm{Im}\bigg[\int_{-\infty}^{0}\frac{d\eta}{\eta^{4}}\left(\pi_k^{(1)*}(0)s_k^{(1)}(\eta) + \pi_k^{(2)*}(0)s_k^{(2)}(\eta) \right)^{2}\cr
&\ \ \ \ \ \ \ \ \ \ \ \ \ \ \ \ \ \ \ \ \ \ \ \ \ \ \ \ \ \ \ \ \ \ \ \ \ \ \ \ \times\left(\beta_{-}(-c\eta)^{\alpha_{-}} + \beta_{+}(-c\eta)^{\alpha_{+}} + \beta_2(-c \eta)^2+\dots\right)\bigg]\nonumber\\
&=V_S'''H^{-4}k^3c^{-3}\textrm{Im}\left[c^{\alpha_{-}}\lambda_{-}(\mu,m) + c^{\alpha_{+}}\lambda_{+}(\mu,m)+c^2\lambda_2(\mu,m)\right]
\end{align} 
where $\beta_{-} = \pi_k^{(1)*}(0)b^{(1)}_{-} +  \pi_k^{(2)*}(0)b^{(2)}_{-}$, $\beta_{+} = \pi_k^{(1)*}(0)b^{(1)}_{+} +  \pi_k^{(2)*}(0)b^{(2)}_{+}$, $\beta_2=\pi_k^{(1)*}(0)b^{(1)}_2+\pi_k^{(2)*}(0)b^{(2)}_2$ and
\begin{align}
\label{lambda- and lambda+}
\lambda_{-}(\mu,m) &=-2 \beta_{-}\int_{-\infty}^{0}\frac{d\eta}{\eta^{4}}\left(\pi_k^{(1)*}(0)s_k^{(1)}(\eta) + \pi_k^{(2)*}(0)s_k^{(2)}(\eta) \right)^{2}(-\eta)^{\alpha_{-}}\cr
\lambda_{+}(\mu,m)  &=-2\beta _{+}\int_{-\infty}^{0}\frac{d\eta}{\eta^{4}}\left(\pi_k^{(1)*}(0)s_k^{(1)}(\eta) + \pi_k^{(2)*}(0)s_k^{(2)}(\eta) \right)^{2}(-\eta)^{\alpha_{+}}\cr
\lambda_{2}(\mu,m)  &=-2\beta _2\int_{-\infty}^{0}\frac{d\eta}{\eta^{4}}\left(\pi_k^{(1)*}(0)s_k^{(1)}(\eta) + \pi_k^{(2)*}(0)s_k^{(2)}(\eta) \right)^{2}(-\eta)^{2}.
\end{align} 
We can compute $\beta_{-}$, $\beta_{+}$, and $\beta_2$ by fitting the numerical mode functions  $s_k^{(i)}(\eta)$ to their power series expansions at small $-\eta$ and extracting $b^{(i)}_{\pm}$, $b^{(i)}_2$ from the fits.  The integrals in (\ref{lambda- and lambda+}) can be computed using the same Wick rotation technique used to compute $B_\pi^\text{equil}$. Then, rearranging (\ref{squeezed integrand leading c behavior}) gives
\begin{align}
\label{squeezed lambda im limit}
&B_\pi^\text{sq} = V_S''' H^{-4}k^3c^{-3/2}\\
&\ \ \ \ \ \ \ \ \ \ \times\left(\textrm{Im}\left[\lambda_{+} + \lambda_{-}\right]\cos\left(\textrm{log}(c){\rm Im}\left[ \alpha_+\right] \right) + \textrm{Re}\left[\lambda_{+} - \lambda_{-}\right] \sin \left( \textrm{log}(c) {\rm Im}\left[\alpha_+\right]\right)+c^{1/2}\textrm{Im}\left[\lambda_2\right] \right)\nonumber
\end{align}
We plot $\textrm{Im}\left[\lambda_++\lambda_-\right]$ in figure~\ref{fig:squeezedcos}. The sine term is usually smaller and so we have not displayed it in a figure.  Equation (\ref{squeezed lambda im limit}) shows that the squeezed limit of the three-point function oscillates logarithmically as a function of $c$.  This behavior is illustrated in figure \ref{fig:squeezedoscillations}. Note that that the dependence of ${\rm Im} \left[\alpha_+\right]=\sqrt{m^2/H^2+ \mu^2/H^2-9/4}$ on $\mu$ has an important effect on the oscillations.  This impacts the two point function of biased objects, see for example~\cite{Gleyzes:2016tdh}.

The oscillatory terms in eq.~(\ref{squeezed lambda im limit}) are enhanced by a factor of $c^{-1/2}$, but are suppressed in the large $\mu$ limit.

\section{Calculating non-gaussianity in the effective theory}
\label{sec:effetivetheory}

\subsection{Brief review of the effective theory for large $\mu$}

In this subsection we begin with a brief review the effective theory approach to the case when  $\mu/H$ is large. In terms of $\pi$ and ${s}$ the Lagrange density is
\bea\label{eq:3pt1}
{\cal L} &=& \frac{1}{2H^2\tau^2}\left[ ({\partial_\tau \pi})^2 - (\nabla\pi)^2 + ({\partial_\tau s} )^2 - (\nabla s)^2 - \frac{m^2}{H^2} \frac{s^2}{\tau^2} - \frac{2 \mu}{H \tau} s {{\partial_\tau \pi}}\right] \nn
&&+ \frac{1}{H^2 \tau^2} \frac{s}{\Lambda} \left[ ({\partial_\tau \pi})^2 - (\nabla\pi)^2 \right] - \frac{1}{H^4\tau^4} \frac{V_S''' s^3}{3!} \nn
\eea

As discussed in Sec.~\ref{Free Field Theory in Flat Space-time}, in flat space-time with  large mixing $\mu$ there is a very massive mode and a massless mode. When $\mu \gg H$ and $ k / a < \mu$, one may integrate out the heavy mode to get an effective theory just involving $\pi$ which can be used to calculate  curvature perturbations.  As discussed in Sec.~\ref{sec:freetheory}, for that purpose the ${(\partial_\tau s)}^2$ and $(\partial_\tau \pi)^2$ terms in eq.~(\ref{eq:3pt1}) can be neglected. Since we assume $m \sim {\cal O}(H)$ or smaller $m$ can also be neglected in eq.~(\ref{eq:3pt1}). With these approximations the equation of motion for $s$ becomes
\beq
0 = \frac{\delta{\cal L}}{\delta{s}} = \frac{1}{H^2 \tau^2} \left[ \nabla^2 s - \frac{\mu\partial_\tau \pi}{H\tau} - \frac{1}{\Lambda}  (\nabla\pi)^2 - \frac{V_S''' s^2}{2 H^2 \tau^2}\right] \ .
\eeq
Up the second order in $\pi$,  the solution for ${s}$ is
\beq
s = \frac{\mu}{H\tau} \frac{1}{\nabla^2} \partial_\tau \pi - \frac{1}{\Lambda} \frac{1}{\nabla^2 }  (\nabla\pi)^2
+ \frac{V_S'''}{2H^2\tau^2} \frac{\mu^2}{H^2\tau^2} \frac{1}{\nabla^2 } \left[ \frac{1}{\nabla^2 } \partial_\tau \pi\right]^2
\eeq
Putting this solution back into eq.~(\ref{eq:3pt1}), the quadratic and cubic terms of the effective Lagrangian of $\pi$ can be written as
\beq
\label{eq:leff2}
{\cal L}^{(2)}_{\rm eff} = -\frac{1}{2H^2 \tau^2} \left[(\nabla\pi)^2 + \frac{\mu^2}{H^2\tau^2} (\partial_\tau \pi) \nabla^{-2} \partial_\tau \pi \right] 
\eeq 
and
\bea\label{eq:leff3}
{\cal L}^{(3)}_{\rm eff} = -\frac{\mu}{\Lambda} \frac{1}{H^3\tau^3} \left[\nabla^{-2} \partial_\tau \pi\right] \left[ (\nabla\pi)^2\right] - \frac{\mu^3}{H^7\tau^7} \frac{V_S'''}{3!} \left[\nabla^{-2}\partial_\tau \pi\right]^3.
\eea
Quantizing the free field part of this effective theory we write for the field operator,
\begin{equation}
 \pi({\bf x},\tau)=\int {d^3 k \over (2 \pi)^3} \left( a({\bf k})  \pi_{k}(\eta) e^{i{\bf k}\cdot {\bf x}} + a^{\dagger}({\bf k})  \pi_{k}(\eta)^{*} e^{-i{\bf k}\cdot {\bf x}} \right).
 \end{equation}
The mode function $\pi_{k}(\eta)$  satisfies the classical equation of motion,
\beq
\label{eq:eqmotion}
\frac{\mu^2}{H^2} \frac{d}{d\eta} \left(\frac{1}{\eta^4} \frac{d\pi_k}{d\eta}\right) + \frac{\pi_k}{\eta^2} = 0 \ .
\eeq
%
which can be solved analytically for the mode function $\pi_k(\eta)$. The normalization of $\pi_k(\eta)$ is determined by the canonical commutation relations. This yields,
%
\beq\label{eq:varphik}
\pi_k(\eta) =  \left(\frac{2\pi^2 \mu}{H}\right)^{1/4} \frac{H}{(2k^3)^{1/2}} \left(\frac{\eta^2 H}{2\mu}\right)^{5/4} H_{5/4}^{(1)} \left(\frac{\eta^2 H}{2\mu}\right)  \ .
\eeq

The power spectrum of the curvature perturbation is
\beq
\label{eq:hurray}
{\cal P}_\zeta = \frac{H^2 }{\dot\phi_0^2} |\pi_k(\eta)|^2_{|\eta| \ll \sqrt{\mu/H}} = \frac{H^4}{\dot\phi_0^2} \left(\frac{1}{2k^3}\right) \left[\frac{16\pi}{\Gamma^2(-1/4)}\left(\frac{\mu}{H}\right)^{1/2}\right] \ .
\eeq
This result was originally derived in Ref.~\cite{Baumann:2011su,Gwyn:2012mw}.

The plot of ${\cal P}_\zeta$ as a function of $\mu$ was shown in Fig.~\ref{fig:funcmu}. The result from the effective theory is shown by the black dashed line. One can see that for $\mu > 10 H$ the result from the effective theory agrees with the numerical result.


\subsection{Non-Gaussianity of equilateral configuration}

The three-point function $B_{\zeta}({\bf  k}_1, {\bf  k}_2, {\bf  k}_3)$ of the curvature perturbation is defined in (\ref{bizeta}). Following standard steps and using the explicit expression of $\pi_k$ in (\ref{eq:varphik}) for the equilateral configuration ($|{\bf k}_1| = |{\bf k}_2| = |{\bf k}_3| =  k$), we have
\beq
B^{\rm equil}_{\zeta}(k) = -\frac{6\mu}{\Lambda} \frac{H^6}{\dot\phi_0^3 k^6} \frac{2^{5/4}\pi^3}{\Gamma^3(-1/4)} {\cal B}_1 - \frac{{V_S'''}}{H} \frac{H^6}{\dot\phi_0^3 k^6} \frac{2^{9/4} \pi^3}{\Gamma^3(-1/4)} {\cal B}_2 \ ,
\eeq
where
\bea
{\cal B}_1 &=& {\rm Re} \int_0^{\infty} dx x^{5/4} \left[ H_{5/4}^{(1)}(x) \right]^{3} \simeq -0.94 \nn
{\cal B}_2 &=& {\rm Re} \int_0^\infty dx x^{-5/2} \left[\frac{d}{dx}\left(x^{5/4} H^{(1)}_{5/4}(x)\right)\right]^3 \simeq -0.09.
\eea
As previously discussed we take
 \bea
 f_{\rm NL}^{\rm equil} &\simeq& \frac{5}{18}\times \frac{B^{\rm equil}_{\zeta}(k)}{({\cal P}_{\zeta}(k))^2} 
 = -\frac{5}{3}\times 2^{-23/4}\pi\Gamma(-1/4)\left[{\cal B}_1 \frac{\mu}{H} + \frac{2}{3} {\cal B}_2 \frac{{V_S'''}}{\mu} \frac{\dot\phi_0}{H^2}\right] \nn&\simeq& -0.45 \times\frac{\mu}{H} - 0.03 \times \frac{{V_S'''}}{\mu}\frac{\dot\phi_0}{H^2} \ .
 \eea
 The factor $\dot\phi_0/H$ can be calculated in terms of the density perturbation and $\mu/H$ using (\ref{eq:DeltaS}). Using the measured value of $\Delta_\zeta$ we have that
 \beq
f_{\rm NL}^{\rm equil} \simeq -0.45\times \frac{\mu}{H} - 140\times \frac{{V_S'''}}{H} \left(\frac{H}{\mu}\right)^{3/4}.
\eeq
 In addition, using the Planck data, the $2\sigma$ constraint on $\mu$ is estimated to be\footnote{Here we have neglected the $V_S'''$ term. It is of course possible that there are cancelations between the contribution proportional to $\mu$ and that proportional to $V'''$ which would relax the bound on $\mu$.}
\beq
\mu/H < 200 \ .
\eeq

\subsection{Non-gaussianity of squeezed configuration}

For the squeezed configuration we consider $k = k_1 \simeq k_2 \gg k_3 = c k$. Taking the contribution from the $1/\Lambda$ term in the interaction Lagrange density we have
\bea
B_{\zeta}^{\rm sq}(k,c) = - \frac{4 H^6}{\dot\phi_0^3} \frac{\mu}{\Lambda} \frac{1}{c^3 k^6} \frac{\pi^3 2^{9/4}}{\Gamma^{3}(-1/4) }({\cal B}_3+{\cal B}_4) \ ,
\eea
where
\begin{align}
{\cal B}_3  &= {\rm Re}~\int_0^\infty dx~ x[H^{(1)}_{5/4}(x)]^2 \left.\frac{d}{dy}\left[y^{5/4}H^{(1)}_{5/4}(y)\right]\right|_{y\rightarrow c^2 x} \\
{\cal B}_4  &= {\rm Re}~\int_0^\infty dx~ xH^{(1)}_{1/4}(x)H^{(1)}_{5/4}(x) \left.\left[c^2y^{5/4}H^{(1)}_{5/4}(y)\right]\right|_{y\rightarrow c^2 x} \ .
\end{align}
Note that $H^{(1)}_{5/4}(x)$ and $H^{(1)}_{1/4}(x)$ oscillate rapidly when $x>1$. Therefore, the integral is mainly supported in the region $x < 1$, which means $c^2 x \ll 1$. Around $y=0$ we have
\beq
y^{5/4}H^{(1)}_{5/4}(y) = -\frac{2^{5/4}i}{\pi}\Gamma(5/4) -\frac{2^{5/4}i}{\pi}\Gamma(5/4)y^2 + {\rm higher~orders} \ ,
\eeq
which implies that 
\beq
\left.\frac{d}{dy}\left[y^{5/4}H^{(1)}_{5/4}(y)\right]\right|_{y\rightarrow c^2 x} = -\frac{2^{9/4}i c^2}{\pi}\Gamma(5/4) x + {\rm higher~orders} \ .
\eeq
For $c\ll1$, ${\cal B}_3$ and ${\cal B}_4$ go like $c^2$ and we have that in the squeezed limit $B_{\zeta}^\text{sq} \sim c^{-1}$.  Even though this contribution is enhanced by a power of $1/c$, it is still suppressed compared to what local non-gaussianity would give which is proportional to ${\cal P}_\zeta(k_1) {\cal P}_\zeta(k_2)  + {\cal P}_\zeta(k_2) {\cal P}_\zeta(k_3) + {\cal P}_\zeta(k_3) {\cal P}_\zeta(k_1)\sim c^{-3}$.  This $c^{-1}$ behavior in the squeezed limit is also seen in equilateral non-Gaussianity.

For the contribution proportional to $V_S'''$ we find
\beq
B_{\zeta}^{\text{sq}}(k,c) = - {V_S''' \over H}\frac{H^6}{\dot\phi_0^3} \frac{1}{c^3 k^6} \frac{\pi^3 2^{9/4}}{\Gamma^{3}(-1/4)} {\cal B}_5 \ ,
\eeq
where in this case
\beq
{\cal B}_5 = {\rm Re}~\int_0^\infty dx~  \left. \left [{ H^{(1)}_{1/4}(x)}\right]^2\left[ {d(y^{5/4}H^{(1)}_{5/4}(y)) \over dy}\right]  \right|_{y\rightarrow c^2 x} 
\eeq
Therefore, the $V_S'''$ interaction also gives a $c^{-1}$ contribution to $B_\zeta^\text{sq}$.

\section{Concluding Remarks}
We studied a simple quasi-single field inflation model where the inflaton couples to another scalar field $S$.  The model contains an unusual mixing term between the inflaton and the new scalar characterized by a dimensionful parameter $\mu$. It has been extensively studied in the literature using perturbation theory in the region where the parameter $\mu/H$ is small and using an effective field theory approach in the region of large $\mu/H$. It has also been studied using numerical methods in other regions of parameter space.  When the mass parameter $m$ of the additional scalar field is zero perturbation theory diverges. 

We numerically calculated the power spectrum and the bispectrum of the curvature perturbations when $\mu$ and the mass  $m$ satisfy $(\mu/H)^{2}+(m/H)^{2} > 9/4$ with $m \sim{\cal O}(H)$ or smaller.  In much of this region, perturbation theory and the effective field theory approach are not applicable.   We found that typically the effective field theory approach is valid for $\mu/H >10$. The numerical approach is non-perturbative in $\mu/H$ and there are no divergences at $m=0$. This occurs because the heavy mode has mass $\sqrt{m^2+\mu^2}$ which does not vanish as $m \rightarrow 0$.

In the case where the inflaton potential is $m_{\phi}^2 \phi^2/2$, we derived  constraints on the parameters $m$ and $\mu$ from $n_S$ and $r$ for $N_{\rm cmb}=50$ and $N_{\rm cmb}=60$. Larger values of $\mu$ make this inflaton potential more compatible with the data.



We computed the contributions from the $\partial \pi \partial \pi s$ and the $s^3$ interactions to the equilateral limit of the bispectrum of the curvature perturbations numerically and compared it with the results from the effective theory.  Using these results and the Planck bounds on $f_{NL}$ we derived upper bounds on $V_S'''$ and $\mu$. 

We also analyzed the squeezed limit of the bispectrum, showing that in this model it is much smaller than for local non-gaussianity.  The contribution to the squeezed bispectrum proportional to $V_S'''$ exhibits interesting oscillatory behavior as a function of the ratio of the small momenta to the larger one.\footnote{This oscillatory behavior was previously noted using perturbation theory for the contribution of the $\partial\pi\partial\pi s$ interaction to the bispectrum and trispectrum~\cite{Arkani-Hamed:2015bza}.} We noted that the oscillation wavelength has $\mu$ dependence that is not evident in perturbation theory.  This behavior could potentially be observed in future experiments.

For small $\mu$ and $m$, there are potentially interesting observational consequences of the behavior of the four point function on the wavevectors that characterize its shape. We will present results on this in a further publication.

\label{sec:conclusion}

\acknowledgments
HA would like to thank Asimina Arvanitaki, Cliff Burgess and Yi Wang for useful comments and discussions. This work was supported by the DOE Grant DE-SC0011632. We are also grateful for the support provided by the Walter
Burke Institute for Theoretical Physics.

\end{document}